# Comparing Approaches to Treatment Effect Estimation for Subgroups in Clinical Trials

Marius Thomas*
Novartis Pharma AG, Basel, Switzerland
and
Björn Bornkamp
Novartis Pharma AG, Basel, Switzerland

**Abstract**

Identifying subgroups, which respond differently to a treatment, both in terms of efficacy and safety, is an important part of drug development. A well-known challenge in exploratory subgroup analyses is the small sample size in the considered subgroups, which is usually too low to allow for definite comparisons. In early phase trials this problem is further exaggerated, because limited or no clinical prior information on the drug and plausible subgroups is available. We evaluate novel strategies for treatment effect estimation in these settings in a simulation study motivated by real clinical trial situations. We compare several approaches to estimate treatment effects for selected subgroups, employing model averaging, resampling and Lasso regression methods. Two subgroup identification approaches are employed, one based on categorization of covariates and the other based on splines. Our results show that naive estimation of the treatment effect, which ignores that a selection has taken place, leads to bias and overoptimistic conclusions. For the considered simulation scenarios virtually all evaluated novel methods provide more adequate estimates of the treatment effect for selected subgroups, in terms of bias, MSE and confidence interval coverage.

*This work was supported by funding from the European Union's Horizon 2020 research and innovation programme under the Marie Sklodowska-Curie grant agreement No 633567 and by the Swiss State Secretariat for Education, Research and Innovation (SERI) under contract number 999754557. The opinions expressed and arguments employed herein do not necessarily reflect the official views of the Swiss Government.

# 1 Introduction

Identifying subgroups, which respond differently to treatment, both in terms of efficacy, as well as safety, is an important part of drug development, see Ruberg & Shen (2015) for an overview. Consequently subgroup analyses are routinely conducted in different forms throughout clinical drug development. Methodological differences in the way subgroup analyses are being conducted exist, depending on when and in what setting they are used.

When there are claims to be made about subgroups in a confirmatory setting in Phase III trials, a multiple testing strategy is set up and separate hypotheses are used to test the treatment effect in the subgroup and the overall population (or the subgroup complement), see for example Bretz et al. (2009) or Glimm & Di Scala (2015) for considerations from a multiple testing perspective.

In case a drug gets approved, additional post-hoc analyses are performed to investigate the consistency of the treatment effect, for example in different countries, or according to other demographic patient characteristics. Alosh et al. (2015) provide an overview of these analyses from a regulatory perspective.

In this article we will focus on exploratory analyses of subgroups in clinical trials, as they are performed for example in early phase clinical trials. The aim in this setting is to investigate, whether patients with different baseline characteristics, such as demographic, disease-specific or genetic covariates respond differently to treatment. Usually a moderate number of covariates is chosen, based on biological, clinical and pharmacological prior information and then analysed. If a subgroup is identified in these analyses, further development of the new drug could then specifically be targeted on the identified subgroup or future trials could be planned using an enrichment strategy, recruiting more patients from the identified subgroup.

This is more challenging than the confirmatory setting, as the sample size in the considered subgroups in these trials is usually too small to identify small (but possibly still relevant) differential treatment effects reliably. In addition multiple groups are being looked at, which leads to a multiplicity/selection problem, in the sense that even if there is no differential effect, the search/selection itself will lead to finding a subgroup with higher observed treatment effect, just by chance (see for example Brookes et al. (2004), Wang et al. (2007), Fleming (2010), Ruberg & Shen (2015)).

Standard medical statistics practice is to perform subgroup analyses in these settings by adding the subgroup covariate (usually as a categorical variable) one at a time in the model as a main effect and as interaction with the treatment. Then the effect of the treatment by covariate interaction is observed to choose a model/subgroup with the most promising treatment effect (*i.e.*, where the difference between subgroup and complement is largest).

The main purpose of this article is to compare novel strategies to improve this practice in realistic simulation scenarios. In particular we evaluate treatment effect estimation after selection in terms of bias, precision and confidence interval coverage, but we also evaluate performance characterictics of our subgroup identification methods.

There exists some work on treatment effect estimation in subgroup analysis, but still the literature is relatively scarce, when it comes to the situation of multiple overlapping subgroups. In the situation of non-overlapping subgroups, hierarchical modelling has become one standard approach to deal with exaggerated treatment effect estimates by a shrinkage towards the overall mean (see for example Jones et al. (2011), Varadhan & Wang (2016)). When it comes to overlapping subgroups Sun & Bull (2005) discuss resampling approaches in the context of genome-wide association studies, a situation which is very similar to the one discussed here. Foster et al. (2011) compare several different ways to estimate the treatment effect in an identified subgroup based on resampling. Rosenkranz (2016) also discusses resampling corrections of treatment effects after a selection has been performed. Berger et al. (2014) discuss Bayesian model averaging to obtain an implicit adjustment for the selection. While not explicitly motivated in the context of subgroup analysis, the Lasso approach proposed by Tibshirani (1996), can also be used to shrink treatment effect estimates in individual subgroups. In this paper the performance of a variant of the model averaging approach (based on Bornkamp et al. (2015)) as well as resampling methods and the Lasso approach are evaluated to obtain corrected treatment effect estimates for a selected subgroup.

In terms of subgroup identification the categorization of continuous covariates has been widely criticized (see among others Weinberg (1995) or Royston et al. (2006)), but is still routinely performed in medical practice. For this purpose, we compare two subgroup identification strategies, one based on categorizing continuous covariates, one based on including continuous covariate effects.

In recent years a number of novel methods for subgroup identification have been developed based on recursive partitioning (see among many others Su et al. (2009), Lipkovich et al. (2011), Foster et al. (2011), Berger et al. (2014), Loh et al. (2015)). As their name suggests, these methods are also based on categorization of covariates to identify subgroups, but they also allow subgroups to be defined in terms of multiple covariates. In this article we will only be concerned with subgroups defined by one covariate, even though most methods presented can be applied with modifications also in situations, when more than one covariate defines the subgroup.

The outline of this paper is as follows. In Section 2 we will present the utilized models and different treatment effect estimates. In Section 3 the setup for the simulation study is presented. In Sections 4 and 5 we will present the simulation results. Section 6 concludes the paper with a short discussion.

## 2 Statistical Methodology

In this section we will present the considered statistical setting, utilized models and different treatment effect estimation methods.

### 2.1 General Setting and Notation

Consider a clinical trial with a treatment and a control arm, $n$ patients in total and a continuous outcome $y$ measured for each patient. Without loss of generality it is assumed that higher values of the outcome variable are preferable with regards to the patient's condition. We assume that this outcome variable is normally distributed, which is the practically most relevant case, but extensions of the models described here to binary, count or time-to-event data should be possible as well.

Furthermore we assume there are $K$ continuous covariates, so that for patient $i$ the $k-$th covariate value is $x_{ik}$, $k = 1, \ldots, K$ and $i = 1, \ldots, n$.

The following linear models are considered for subgroup identification and subgroup effect inference. We denote the total number of models by $P$, where $P$ can be larger than $K$, so there can be more models than covariates. Then for $p = 1, ..., P$ each model is of the general form

$$M_p : y_i \sim N(\mu_i^{(p)}, \sigma_p^2), i = 1, ..., n,$$
$$\text{where } \mu_i^{(p)} = \beta_0^{(p)} + h^{(p)}(x_{i.}) + (\beta_1^{(p)} + g^{(p)}(x_{i.}))T_i \qquad (1)$$

In this model $T_i$ denotes whether patient $i$ is in the treatment ($T_i = 1$) or the control arm ($T_i = 0$), $x_{i.} = (x_{i1}, ..., x_{iK})'$ contains the covariates for patient $i$. $h$ and $g$ are functions of the covariates, mapping from $\mathbb{R}^K \to \mathbb{R}^1$. In many of the cases we consider here, these function transform only one of the $K$ covariates, so that each of the $P$ models only consider one covariate at a time. The function $h(x)$ represents prognostic (modifying the response independent of treatment) and the function $g(x)$ predictive effects (modifying the response to treatment) of a covariate.

In what follows two different choices for $h$ and $g$ are discussed.

## 2.2 Choosing the Functions $h$ and $g$

### 2.2.1 Step Functions

A standard approach in terms of subgroup analyses is to choose the functions $g$ and $h$ proportional to step functions of the form $\mathbb{1}_{(c,\infty)}(x)$, where $c$ is a cut-off and $\mathbb{1}$ is the indicator function. This cut-off splits the trial population into two parts with consequently different response. Often $c$ is chosen as a quantile of the corresponding covariate vector. In the following the three quartiles of each covariate are used as cut-offs. Using the quartiles still allows for a certain amount of flexibility, but the overall population is not partitioned into too small parts. In detail this approach results in the following functions $g$ and $h$:

$$h^{(p)}(x_{i.}) = \beta_2 s_i^{(p)} \text{ and } g^{(p)}(x_{i.}) = \beta_3 s_i^{(p)},$$

$$\text{where } s_i^{(p)} := \begin{cases} \mathbb{1}_{(q_{0.25}(x_{.1}),\infty)}(x_{i1}) & \text{for } p = 1 \\ \mathbb{1}_{(q_{0.5}(x_{.1}),\infty)}(x_{i1}) & \text{for } p = 2 \\ \mathbb{1}_{(q_{0.75}(x_{.1}),\infty)}(x_{i1}) & \text{for } p = 3 \\ \mathbb{1}_{(q_{0.25}(x_{.2}),\infty)}(x_{i2}) & \text{for } p = 4 \\ \vdots & \\ \mathbb{1}_{(q_{0.5}(x_{.K}),\infty)}(x_{iK}) & \text{for } p = P - 1 \\ \mathbb{1}_{(q_{0.75}(x_{.K}),\infty)}(x_{iK}) & \text{for } p = P, \end{cases}$$

and $q_p(x_{.k})$ denotes the $p-$quantile of the $k$th covariate.

Each of the vectors $s^{(p)} = (s_1^{(p)}, ..., s_n^{(p)})'$ is thus a binary vector, denoting, whether a patients belongs to a subgroup (*i.e.* the corresponding covariate is larger than the cut-off) or not. So each covariate defines three different partitions, which split the trial population into subgroup and complement. With this approach for transforming covariates there are thus $P = 3 \cdot K$ candidate models of the form

$$\begin{aligned} M_p &: y_i \sim N(\mu_i^{(p)}, \sigma_p^2), i = 1, ..., n \\ \text{where } \mu_i^{(p)} &= \beta_0^{(p)} + \beta_2^{(p)} s_i^{(p)} + (\beta_1^{(p)} + \beta_3^{(p)} s_i^{(p)})T_i \end{aligned} \quad (2)$$

### 2.2.2 Spline Modelling

Instead of using step functions one can also employ splines to model the effect of covariates, here natural cubic regression splines will be used (see for example (Hastie et al. 2009, Chapter 9)). We use

$$h^{(p)}(x_{i.}) = \sum_{m=1}^{L} \beta_{m+1}^{(p)} \cdot b_m(x_{ip})$$

$$g^{(p)}(x_{i.}) = \sum_{m=1}^{L} \beta_{L+m+1}^{(p)} \cdot b_m(x_{ip}),$$

here $b_1, ..., b_L$ are the basis functions of a natural cubic spline for $L$ given knots $\xi_1, ..., \xi_L$. See Appendix A for how the knots were selected, depending on the total sample size.

By using these forms of $g$ and $h$ in model (1) one obtains $P = K$ models in total, one model for each covariate.

$$M_p : y_i \sim N(\mu_i^{(p)}, \sigma_p^2), i = 1, ..., n,$$
$$\text{where } \mu_i^{(p)} = \beta_0^{(p)} + \sum_{m=1}^{L} \beta_{m+1}^{(p)} \cdot b_m(x_{ip}) + (\beta_1^{(p)} + \sum_{m=1}^{L} \beta_{L+m+1}^{(p)} \cdot b_m(x_{ip}))T_i. \quad (3)$$

Transformations of covariates are often used in practice to obtain more uniformly distributed covariates (*e.g.* using a log-transformation for positive measurements). We use a rank transformation, *i.e.*, instead of using $x_{ik}$ per patient, $r_{ik} = rank(x_{ik})/n$ will be used as input to the spline bases, where $rank(x_{ik})$ gives the rank of $x_{ik}$ within all values $x_{1k}, \ldots, x_{nk}$ of covariate $k$. Of course all modelling results can be backtransformed using the quantile function, if desired. Using this approach the flexibility of the spline will be higher where the covariate is densely sampled and the flexibility of the spline is reduced where there are only few data. In our experience this approach improves performance of the method by reducing the variability of the spline estimate close to the boundaries of the covariate space.

### 2.2.3 Obtaining Treatment Effects

We now describe how a treatment effect estimate for an identified subgroup can be obtained from the two models above. Assume a subgroup of patients has been identified and let $S$ denote the set of indices (from $\{1, \ldots, n\}$) corresponding to the patients in this subgroup. Denote the treatment effect in $S$ by $\Delta(S)$. An estimate of $\Delta(S)$ can be derived by predicting the treatment effect for each patient in the subgroup and averaging over all patients in the subgroup. Using any model $M_p$ from (1) we predict the treatment effect for patient $i$ under $M_p$ as

$$\hat{\delta}_i^{(p)} = \hat{\mu}_{i|T_i=1}^{(p)} - \hat{\mu}_{i|T_i=0}^{(p)} = \hat{\beta}_1^{(p)} + \hat{g}^{(p)}(x_{i.}). \quad (4)$$

The estimated treatment effect for a subgroup $S$, under model $M_p$, is then given by

$$\hat{\Delta}^{(p)}(S) = \frac{1}{|S|} \sum_{i \in S} \hat{\delta}_i^{(p)}. \quad (5)$$

For the model (2) this reduces to

$$\hat{\Delta}^{(p)}(S) = \hat{\beta}_1^{(p)} + w\hat{\beta}_3^{(p)}, \tag{6}$$

where $w = |S \cap S^{(p)}|/|S|$ and $S^{(p)} = \{i \in \{1, ..., n\}|s_i^{(p)} = 1\}$ is the subgroup investigated in model $M_p$. The value $w \in [0, 1]$ is measuring the fraction of patients of subgroup $S$ that are also in subgroup $S^{(p)}$. This simplifies to $\hat{\beta}_1^{(p)} + \hat{\beta}_3^{(p)}$, if $S = S^{(p)}$ and to $\hat{\beta}_1^{(p)}$ if $S$ is the complement of $S^{(p)}$. For model (3) one obtains $\hat{\Delta}^{(p)}(S) = \hat{\beta}_1^{(p)} + \sum_{m=1}^{L}[\hat{\beta}_{L+m+1}\frac{1}{|S|}\sum_{i\in S}b_m(x_{ip})]$ as the treatment effect estimate under a given model $M_p$.

## 2.3 Subgroup Identification

The focus of this work is treatment effect estimation, so for identification of the subgroup we used approaches that we think reflect current practice relatively well. The recursive partitioning methods cited in the introduction would be alternative more elaborate methods for identification of subgroups, that also allow to identify subgroups defined by more than one covariate.

For identifying a subgroup, interaction tests on the predictive terms of the models are employed. We only test the predictive term, because the interest lies in finding subgroups, that have a better treatment effect than the overall population. Prognostic effects occur independently of the treatment and don't influence the treatment effect. For model (2) this leads to a t-test on the null hypothesis $H_0 : \beta_3 = 0$. For the model (3) a likelihood-ratio test is performed comparing the models with and without the predictive term $g^{(p)}(x_{i.})$. In both scenarios these tests are performed for each of the $P$ submodels. Then the model with the highest evidence of a differential effect (as measured by the lowest p-value) is identified as the model for describing a subgroup. When using a step function this approach is equivalent to selecting the subgroup, where the standardized treatment effect difference is maximized.

When using step functions picking a model is equivalent to choosing a subgroup, for splines only the covariate is chosen. In case splines are used, the subgroup definition still needs to be derived. For consistency a similar criterion will be used for splines as for step functions to obtain a cut-off value.

We identified the cut-off, that maximizes the differential treatment effect between the two resulting subgroups. Assume that covariate $p$ has been identified. We then calculate a metric for each cut-off from a prespecified set, that splits the patients into the subgroup $S_1 = \{i \in \{1, ..., n\} | x_{ip} > c\}$ and its complement $S_2 = \{i \in \{1, ..., n\} | x_{ip} \leq c\}$. We use a metric of the form

$$Z(S_{1,c}, S_{2,c}) = \frac{|\hat{\Delta}^{(p)}(S_{1,c}) - \hat{\Delta}^{(p)}(S_{2,c})|}{\sqrt{Var[\hat{\Delta}^{(p)}(S_{1,c}) - \hat{\Delta}^{(p)}(S_{2,c})]}}, \tag{7}$$

and choose the cut-off $c$ which maximizes this from a set of possible cut-offs $c_1, ...., c_s$. In analogy to subgroup identification using step functions we consider the three quartiles as possible cut-offs for the chosen covariate.

With both approaches we technically end up with two possible subgroups, left and right of the identified cut-off. To identify which of those subgroups has the better treatment effects, the naive estimates (see Section 2.4.1) are compared and the subgroup with the bigger estimate is then identified as the best subgroup.

The approach in this section chooses the subgroup for which the evidence of a differential effect between subgroup and complement is largest. An alternative would be to select the subgroup where the estimated treatment effect difference to the complement is largest (*i.e.* ignoring the denominator in Equation (7)). We also considered this approach and will come back to this in the Discussion in Section 6.

## 2.4 Treatment Effect Estimation in Subgroups

In this section the different methods for treatment effect estimation will be presented. Apart from the naive estimation method, all of the following methods try to provide an adjusted estimate that corrects for the bias, which is introduced through the selection of a subgroup. Note that all methods can be used based on both models (step functions and splines) introduced above.

### 2.4.1 Naive Estimates

For the naive estimates we simply estimate the treatment effect as in (5) from the model that corresponds to the selected subgroup (when using (2)) or the covariate identified to define the subgroup (when using (3)). A naive treatment effect estimator for an identified subgroup S defined by model $M_p$ is therefore simply

$$\hat{\Delta}_{naive}(S) := \hat{\Delta}^{(p)}(S), \tag{8}$$

where $S = S^{(p)}$. The variance for these estimates can be derived by observing that this is an average of linear combinations of maximum likelihood parameter estimates, regardless of which modeling approach is used. A $(1-\alpha)$- confidence interval is then given through

$$[\hat{\Delta}_{naive}(S) \pm t_{n-d,1-\frac{\alpha}{2}} \sqrt{Var(\hat{\Delta}_{naive}(S))}], \tag{9}$$

where $t_{n-d,1-\frac{\alpha}{2}}$ is the $(1-\frac{\alpha}{2})$-quantile of the t-distribution with $n-d$ degrees of freedom and d is the number of parameters in the model ($d = 4$ for models (2) and $d = 2L+2$ for models (3)). Note that this approach is problematic as the selection process/model uncertainty is being ignored: Model $M_p$ is utilized as if it was pre-specified and not selected in a data-driven manner.

### 2.4.2 Model Averaging

Subgroup identification can be viewed as a model selection: A set of candidate models (= subgroups) is evaluated and then one is chosen, ignoring the associated model uncertainty. An alternative is to use model averaging see among many others Draper (1995), Raftery (1995). The idea of model averaging in the context of treatment effect estimation for subgroups is that every model $M_p$ can be used to predict a treatment effect for a subgroup $S$. Because most covariates do not interact with the treatment, many of these models will predict a similar treatment effect for the underlying subgroup and complement, which will both be similar to the overall effect. Depending on the weight of the different models, there will hence be more or less shrinkage towards the overall effect for a particular subgroup $S$. Note that the naive estimates can be seen as a special case of this approach, where one model is given a model weight of 1 and all others zero. From a model averaging perspective this naive estimate would thus only be justified if the corresponding model would be much more likely than any of the other considered models.

Using Bayesian model averaging in the context of subgroup analyses was proposed by Berger et al. (2014) with a particular focus on subgroup identification and less on treatment effect estimation. They used a full Bayesian approach, and also performed model averaging over the model with no treatment effect and the model with a homogeneous treatment effect. For the simulations here, we chose to use BIC approximations for the model weights. The BIC turns out to give similar weights to the ones obtained by a full Bayesian approach, based on MCMC sampling, but is much faster to calculate. See Bornkamp et al. (2015) for a more detailed comparison of MCMC versus BIC approximations and inclusion of the models assuming a homogeneous or no treatment effect in a model averaging approach. The BIC approximations for posterior model weights are given by

$$P(M_p|y) \approx P^*(M_p|y) = \frac{\exp(-0.5 \cdot BIC(M_p))}{\sum_{p'=1}^{P} \exp(-0.5 \cdot BIC(M_{p'}))}.$$

Note that all models have the same number of parameters here, so that the penalty terms cancel out and the $AIC$ or just the log-likelihood instead of the $BIC$ would give identical results.

By using the posterior model weights, estimates for the posterior mean and variance of the treatment effect of a subgroup $S$ can be obtained by estimating the treatment effect of this subgroup under all considered models, as described in Section 2.2.3, and then averaging over all models. The model averaging estimator of the treatment effect is then the approximation to the posterior mean,

$$\hat{\Delta}_{ma}(S) := \sum_{p=1}^{P} P^*(M_p|y)\hat{\Delta}^{(p)}(S).$$

The posterior variance can be approximated by

$$\sum_{p=1}^{P}[Var(\hat{\Delta}^{(p)}(S)) + (\hat{\Delta}^{(p)}(S))^2]P^*(M_j|y) - \hat{\Delta}^2_{ma}(S),$$

which can be used to calculate an uncertainty interval based on a normal approximation (see also Raftery (1995)).

### 2.4.3 Resampling

The general idea of the resampling approach is to split the original dataset into an identification/selection sample and an estimation sample, so patients are either in the identification or estimation sample. This allows to estimate the magnitude of over-estimation introduced by the selection. Sun & Bull (2005) compare several different resampling approaches. Two estimators from their paper will be used here. The first tries to estimate the bias due to the selection process directly (similar methods were discussed by Foster et al. (2011) and Rosenkranz (2016)) the other being the 0.632-estimator, originally derived for estimation of the prediction error of classification rules (Efron (1983)), which performed best in the comparisons of Sun & Bull (2005). Additionally we use a third method, which is a resampling version of the model averaging approach and therefore more similar to the approach described in the previous section. For this paper we used bootstrap resampling estimators.

If a subgroup $S$ has been identified, the naive estimate for the treatment effect in that subgroup is $\hat{\Delta}_{naive}(S)$ and is obtained as described in Section 2.4.1. To obtain adjusted resampling estimates, $B$ bootstrap samples of size $n$ are drawn from the original dataset and the identification process is repeated in each of these identification samples. In each of the $B$ steps the treatment effect for the identified subgroup is estimated using the identification sample and the out-of-sample patients in the estimation sample, the difference between these two estimates should give an idea of the bias induced by selection. Note that the identification in each of the $B$ steps might lead to different subgroups than the one identified originally on the whole dataset.

The first estimator tries to estimate the selection bias by comparing the estimates in the identification and estimation samples directly. If $\hat{\Delta}_{naive}^{(I)}(S_I^{(b)})$ denotes the naive estimate from the $b$-th identification sample for the subgroup, $S_I^{(b)}$, identified in the same identification sample and $\hat{\Delta}_{naive}^{(E)}(S_I^{(b)})$ denotes the estimate from the $b$-th estimation sample for the same subgroup. Then the difference between the two gives an estimate of the selection bias. Then the estimator for $\Delta(S)$ is

$$\hat{\Delta}_{rsbias}(S) := \hat{\Delta}_{naive}(S) - \frac{1}{B}\sum_{b=1}^{B}[\hat{\Delta}_{naive}^{(I)}(S_I^{(b)}) - \hat{\Delta}_{naive}^{(E)}(S_I^{(b)})]. \qquad (10)$$

This estimator adjusts the naive estimate by substracting the estimate for the selection bias.

The 0.632-estimator uses the approximation that on average the sample size in the estimation sample in each step is $n_E = 0.368n$, since on average 36.8 % of patients do not appear in the identification sample. The adjusted estimate derived from this combines the naive estimate with the average estimates in the estimation sample so that the adjusted estimate for the originally identified subgroup is

$$\hat{\Delta}_{rs632}(S) := (1 - 0.632)\hat{\Delta}_{naive}(S) + 0.632 \cdot \frac{1}{B} \sum_{b=1}^{B} \hat{\Delta}_{naive}^{(E)}(S_I^{(b)}). \quad (11)$$

For simplicity, we calculated confidence intervals for these two methods using the intervals for the naive estimator, but centering it around the bias-adjusted estimates.

The third resampling estimator is motivated by bootstrap model averaging in the sense of a bagging estimator (Breiman (1996)). For this estimator we use the $BIC$ in each bootstrap sample to identify the model with the best subgroup/covariate. Then the treatment effect for the originally identified subgroup $S$ is estimated under this model. We obtain an estimate for the treatment effect by averaging over the estimates for all bootstrap samples. If $\hat{\Delta}^{(b)}(S)$ therefore denotes the estimate of the treatment effect in $S$ from the model $M_b$, which was identified as the one with the lowest BIC from all candidate models in the $b$-th bootstrap sample, an estimator for the treatment effect in $S$ is given through

$$\hat{\Delta}_{rsma}(S) := \frac{1}{B} \sum_{b=1}^{B} \hat{\Delta}^{(b)}(S). \quad (12)$$

Since we have an estimate of the treatment effect for $S$ in each bootstrap sample we can use the empirical quantiles from these $B$ estimates to calculate a bootstrap confidence interval for the treatment effect in the subgroup.

### 2.4.4 Lasso Regression

The models described in Section 2.2 are all marginal models, in the sense that they only include one covariate/subgroup at a time in each model. An alternative to fitting several marginal models is fitting a single multivariate model, which includes prognostic and predictive effects for all covariates. As it is expected that only a small number of covariates will have an effect on the response and the treatment effect, it makes sense to use penalized regression methods.

Lasso regression as presented by Tibshirani (1996) is maybe the most commonly used penalized regression method. Penalization is achieved by adding a penalty to the log-likelihood function, so instead of maximizing the log-likelihood alone the penalty term $\lambda \sum_{j=1}^{K} |\beta_j|$ is added, which induces shrinkage towards 0, the parameter $\lambda$ controls the shrinkage.

For the Lasso estimates we hence use only one model, including all covariates. In the notation of the general modeling approach (1) we use the terms

$$h(x_{i.}) = \sum_{k=1}^{K} \beta_{k+1} \cdot x_{ik}$$

$$g(x_{i.}) = \sum_{k=1}^{K} \beta_{K+k+1} \cdot x_{ik}.$$

The $R$ package *glmnet* (Friedman et al. (2009)) was used to estimate the Lasso model. We did not add a penalty on the parameter for the overall effect ($\beta_1$ in model (1)), since the treatment effect estimates in identified subgroups should be shrunken towards the overall treatment effect but not below it. Therefore only the parameters in the prognostic and predictive terms $g$ and $h$ were included in the penalty. We chose the shrinkage parameter $\lambda$ that minimised the average cross validation error over 10 10-fold cross validations using the *cv.glmnet* function, since choosing the $\lambda$ on a single 10-fold cross validation was found to be too variable for smaller sample sizes.

Based on this model a predicted treatment effect $\hat{\delta}_i$ for patient $i$ can be obtained as in Equation (4). An estimator for the treatment effect in subgroup $S$ based on the Lasso model is therefore given by the average of the predicted treatment effects of patients in the subgroup as

$$\hat{\Delta}_{las}(S) := \frac{1}{|S|} \sum_{i \in S} \hat{\delta}_i, \tag{13}$$

where $\hat{\delta}_i = \hat{\beta}_1^{(L)} + \sum_{m=1}^{K} \hat{\beta}_{K+m+1}^{(L)} \cdot x_{im}$, and $\hat{\beta}_1^{(L)}$, $\hat{\beta}_{K+m+1}^{(L)}$ are the estimates determined by the Lasso. Note that a few of the $\hat{\beta}_{K+m+1}^{(L)}$ will be estimated to be exactly 0. In the extreme case, where all of the $\hat{\beta}_{K+m+1}^{(L)}$ will be zero, Lasso will predict the same treatment effect for every patient. To come up with a confidence interval again the naive interval is re-centered around this adjusted estimate (as for $\hat{\Delta}_{rs632}$ and $\hat{\Delta}_{rsbias}$).

Note that the Lasso approach is performed here not to identify a subgroup, just to obtain the treatment effect for the already determined subgroup (by either the step function or spline approach). Contrary to the model averaging and resampling approaches it is independent of the models that are used for identification.

# 3 Simulation Setup

In this section we will describe how we set up the simulation study. Two subgroup identification methods were used and six methods for estimating treatment effects (see Table 1 for an overview), giving 12 methods to evaluate in total. We used $R$ (R Core Team (2015)) for simulations. For all scenarios we ran 5000 simulations with $B=100$ bootstrap samples.

Table 1: Estimators used in simulations

| estimation method (estimator) | denoted by |
|:---:|:---:|
| naive ($\hat{\Delta}_{naive}$) | naive |
| model averaging ($\hat{\Delta}_{ma}$) | ma |
| 0.632-estimator ($\hat{\Delta}_{rs632}$) | rs632 |
| resampling with model averaging ($\hat{\Delta}_{rsma}$) | rsma |
| resampling bias adjustment ($\hat{\Delta}_{rsbias}$) | rsbias |
| lasso ($\hat{\Delta}_{las}$) | Lasso |

## 3.1 Data Generation

The covariate vectors $x_{.1}, ..., x_{.K}$ are generated from independent standard normal distributions for each patient. The outcome variable $y_i$ is generated from the covariates using the model

$$y_i = T_i \cdot g(x_{i1}) + \epsilon_i, \ i = 1, ..., n, \tag{14}$$

where the $\epsilon_i$ are i.i.d standard normally distributed. This means that the response mean under the control group is 0, in addition there are no prognostic covariates that have an effect on the outcome independent of the treatment. The treatment effect depends on the first covariate and is described by the function $g$. The model above implies that the overall average treatment effect, which we will denote here by $\Delta(\Omega)$ is given by

$$\Delta(\Omega) = \int_{-\infty}^{\infty} g(x) \cdot f_{N(0,1)}(x) dx,$$

where $f_{N(0,1)}(x)$ is the density of the standard normal distribution.

## 3.2 Simulation Parameters

Table 2 gives an overview of the different parameters varied in the simulation study.

Table 2: Parameter settings varied in the simulations.

| parameter | abbreviation | settings |
|---|---|---|
| sample size | $n$ | 50, 500 |
| number of covariates | $K$ | 5, 10, 30 |
| effect size | delta | large (power=0.9), small (power=0.5), zero |
| treatment effect curve | curve | step function, linear, sigmoidal |

For the sample size $n$ of the trial, we chose to use $n = 50$ and $n = 500$. A sample size of $n = 50$ could be considered representative of an early Phase II trial (a Proof-of-Concept (PoC) trial). These trials are usually the first trials with patients. The aim of a Proof of Concept-trial is to determine, whether there is enough evidence for the treatment effect of a drug to warrant further development. If this is the case, the drug will be developed further. In this scenario interest often also lies on finding potential groups of patients, that respond differently to the drug, in particular if the overall treatment effect is not promising enough. The second sample size of $n = 500$ is more representative of a small Phase III study. As discussed in the introduction, exploratory analyses are also routinely conducted in this later phase of drug development.

The treatment effect in the overall population is another parameter that is varied across the simulations. Three different effect sizes are used in the simulation study. First we consider a "zero" scenario of no overall treatment effect and no subgroup. In the other two scenarios there is an overall treatment effect. The overall effect sizes are based on the power they would give a two-sample t-test (with significance level 5% two-sided) to show a significant treatment effect on the overall population. The two power values used in this simulation study are 0.9 and 0.5, we will in the following denote this by "large" and "small" overall treatment effects. The first scenario covers the situation, where the sample size is sufficient to reliably detect the overall treatment effect. The second scenario covers the situation, where the sample size is too small to detect the overall treatment effect, and only in a subgroup a promising treatment effect exists (but the sample size in the subgroup is too small to detect the effect).

The functional form of the interaction between covariates and treatment, which determines how the treatment effect is distributed across patients, can have an influence on subgroup identification and subgroup effect estimation. Hence in this simulation study three different forms of $g$ are used, modelling the treatment effect depending on the first covariate. The three forms are a step function, a linear function and a curve with a sigmoidal form. The corresponding equations to these forms of g are

$$\begin{aligned} g_{step}(x) &= b_{step} \cdot \mathbb{1}_{(0,\infty)}(x) \\ g_{lin}(x) &= a_{lin} + b_{lin} \cdot x \\ g_{sig}(x) &= a_{sig} + b_{sig} \cdot erf(x) \end{aligned}$$

where $erf(x) = \frac{2}{\sqrt{\pi}} \int_0^x e^{-\tau^2}$ is the error function and the specific intercept and slope parameters, $a$ and $b$ are chosen so that an overall average treatment effect $\Delta(\Omega)$ is achieved, in addition it is required that the average treatment effect for patients i with $x_{i1} > 0$ is given by $2 \cdot \Delta(\Omega)$. This second condition guarantees that there are subgroups with higher than average treatment effects. Given these conditions we can determine $a$ and $b$ for all functions $g$, see Appendix B and Table 3.

Table 3: Overall treatment effects and parameter values for the different treatment effect functions. For $g_{lin}$ and $g_{sig}$ the intercept $a$ is equal to the overall effect.

| settings | overall | step | linear | sigmoidal |
|---|---|---|---|---|
| $n = 50$, power=0.9 | 0.94 | b=1.88 | $b = 1.17$ | b=1.54 |
| $n = 50$, power=0.5 | 0.57 | b=1.14 | $b = 0.71$ | b=0.48 |
| $n = 500$, power=0.9 | 0.29 | b=0.58 | $b = 0.36$ | b=1.07 |
| $n = 500$, power=0.5 | 0.18 | b=0.36 | $b = 0.22$ | b=0.29 |

See Figure 1 for a graphical display of the different functions. For covariate values $x_1 < 0$ patients have a treatment effect smaller than the overall treatment effect in all cases. For the linear and sigmoidal function the response even gets worse than the control response mean (which is 0). This is not unreasonable to assume, when considering that an active control might be used in the trial.

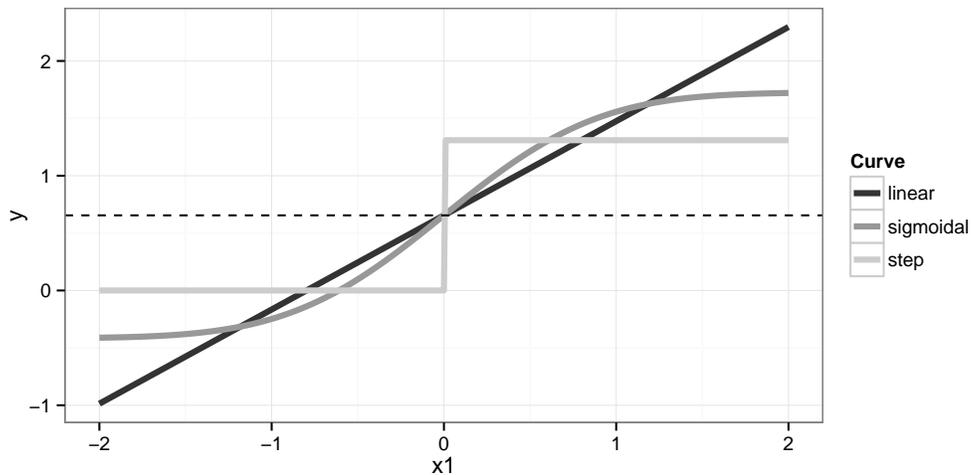

Figure 1: Different types of (non-zero) treatment effect curves used in the simulation study for sample size $n = 100$ and power 0.9, the dashed line shows the corresponding overall treatment effect of 0.65.

Another important parameter is the number of covariates $K$ that are used to identify possible subgroups. We chose $K=5$, 10 and 30. Clearly it is expected that a higher number of covariates makes correct subgroup identification, as well as treatment effect estimation more difficult, since the possibility of chance findings increases.

## 4 Simulation results

We simulated 5000 datasets under each combination of simulation parameters. For each of the datasets we identified the best subgroup with the simple identification strategy described in Section 2.3, choosing the model with the lowest p-value for the interaction test.

We then estimate the treatment effect for the identified subgroup (which is not necessarily the correct one) using all estimators shown in Table 1. The main performance metrics considered are the bias, MSE and the coverage probability of 90% confidence intervals calculated as described in Section 2. We use median biases and MSEs, since the estimates obtained with the spline approach were prone to contain outliers.

We briefly present the differences between the two modeling approaches in subgroup identification performance and then turn to the results for treatment effect estimation, which is the main part of our work.

## 4.1 Performance of Subgroup Identification Methods

Figure 2 summarizes the performance of the two identification approaches in different scenarios for 10 covariates. Results for the remaining two settings for the number of covariates are shown in the supplementary material.

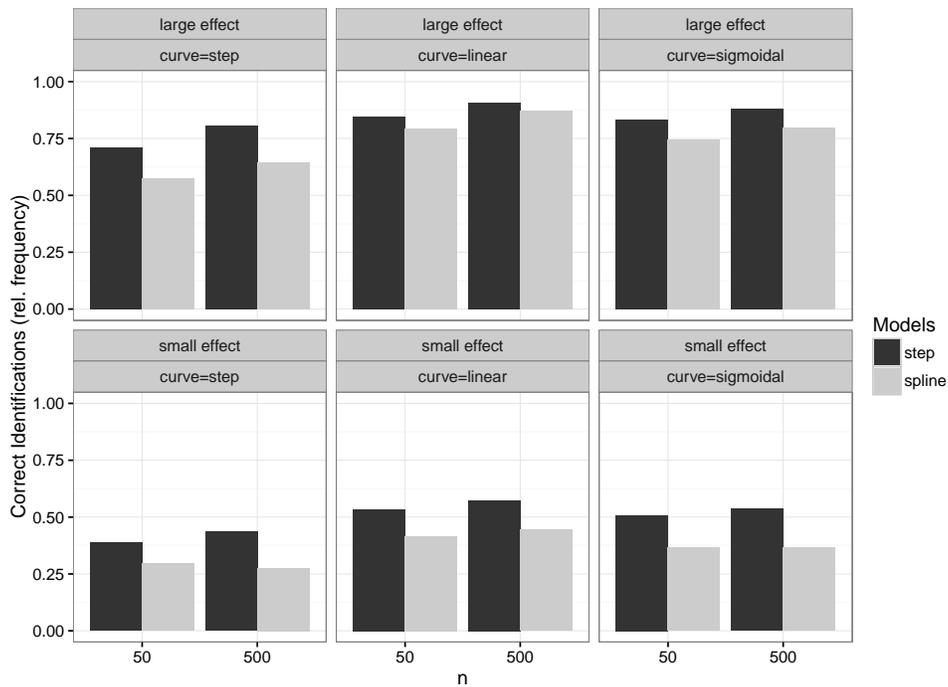

Figure 2: Relative frequency of correct identifications for the spline and step function modeling approaches. An identification is considered as correct, when there is a true interaction between the identified covariate and the treatment. In all depicted scenarios the number of covariates was 10.

The step function approach identifies the correct covariate with a similar or even higher percentage than the spline approach in all scenarios with an existing subgroup. The difference is largest, when the true treatment effect curve is a step function. If the true treatment effect curve is either linear or sigmoidal the difference between the two methods is smaller, as expected as in this case the step functions models are misspecified. It is nevertheless interesting that even in those scenarios step functions outperform the splines, we will return to this point in the discussion.

## 4.2 Simulation Results for Effect Estimation in Subgroups

In this section we only show the bias, MSE and confidence interval coverage for a sample size of $n = 50$ and for a true step function treatment effect curve. Results for the larger sample size and other functional forms of treatment effect curves were similar and are shown in the supplementary material. The results for the bias of the different estimators are shown in Figure 3, MSE results are depicted in Figure 4 and coverage of confidence intervals is depicted in Figure 5.

One general observation is that the biases in Figure 3 tend to get larger the smaller the treatment effect in the true subgroup is. This is because the chances to identify the correct subgroup are naturally higher, if the effect in that subgroup is big (see also Section 4.1). For small effects the found subgroup is more likely to be a chance finding.

The naive estimates overestimate the true treatment effects, especially, when the effect in the true subgroup is only small or there is no subgroup and effect. The bias also increases for a higher number of covariates. Comparing the two modeling approaches, the spline method seems to result in slightly better naive estimates, since the biases are usually a bit smaller.

The ma-, rsma- and rs632-estimators all behave similarly and have smaller biases than the naive estimates. For large effect sizes these three estimators tend to have a slight negative bias, while the bias is positive, but smaller than the naive estimate, when effects are small or zero. An additional observation for the ma-estimator is that the bias for spline models seems to be slightly bigger than for step-function models. An intuitive explanation for this result, is that using model averaging with spline models does not properly account for the uncertainty in the selection of the subgroup cut-off, since cut-offs are determined

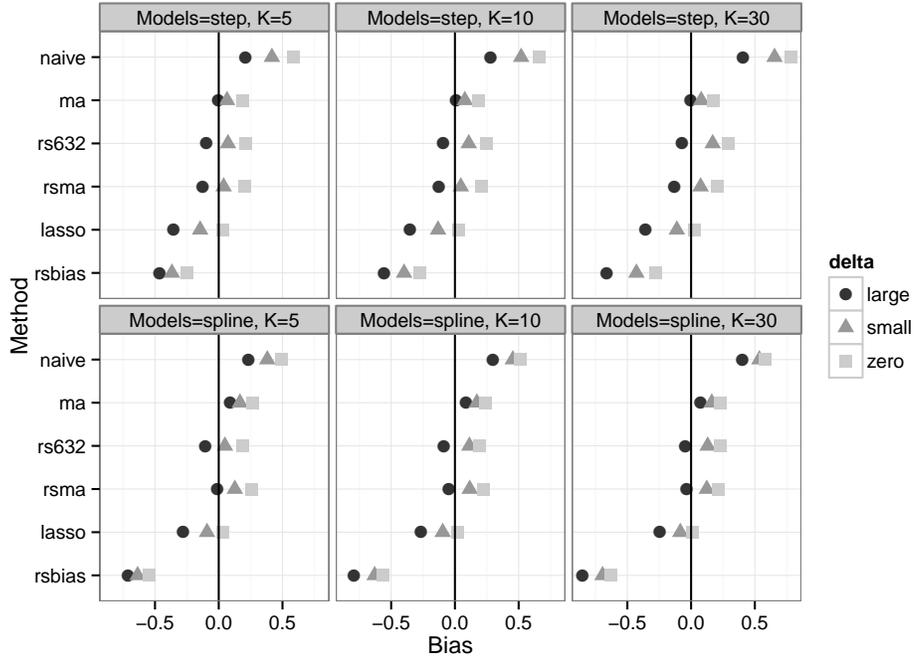

Figure 3: Median bias of estimators for the step function (top row) and spline modeling approaches. Scenarios with differing number of covariates $K$ and effect sizes *delta* in the true subgroup are depicted. For all depicted scenarios $n = 50$ and *curve = step*.

after choosing the best model.

The Lasso estimator has a small bias, when the effect size in the true subgroup is small or there is no subgroup at all. Generally it has a more negative bias than all of the estimators described before for the case of a large effect size.

The rsbias estimator shows very large biases and always underestimates the true effect, even if the true effect is zero, which means that it gives estimates that lie below the overall treatment effect.

For the MSE (Figure 4) the results are similar to the results for the bias. The MSEs for the naive estimates are higher than for most of the other estimators, but are smaller for the spline method than for the step functions. The ma- rsma-, rs632 and Lasso-estimators have smaller MSEs than the naive estimates. When the effect in the subgroup is small or there is no subgroup at all, the Lasso-estimator has the smallest MSE of all estimators evaluated here. The rsbias-estimator again performs poorly and has sometimes worse MSEs than the naive estimates.

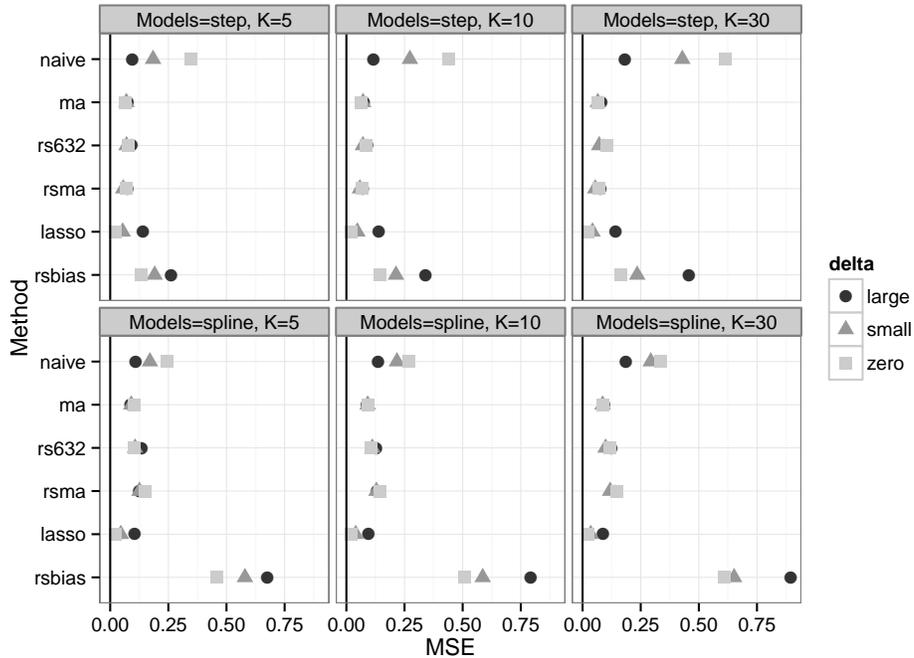

Figure 4: Median MSE of estimators for the step function (top row) and spline modeling approaches. Scenarios with differing number of covariates $K$ and effect sizes *delta* in the true subgroup are depicted. For all depicted scenarios $n = 50$ and $curve = step$.

In terms of confidence interval coverage (Figure 5), the naive estimates are always below the nominal level, sometimes even dropping below 50%, when there is no subgroup and the number of covariates is high. The rsbias estimator is also always below the nominal coverage. The calculated confidence intervals for the remaining estimators generally have good properties (in particular rsma, rs632, and ma) and usually have a coverage around the nominal level.

# 5 Further Simulations Investigating Asymptotic Behaviour

In this section we evaluate the influence of certain simulation parameters on the estimates given by the different methods, going beyond the scenarios shown in Table 2, in particular

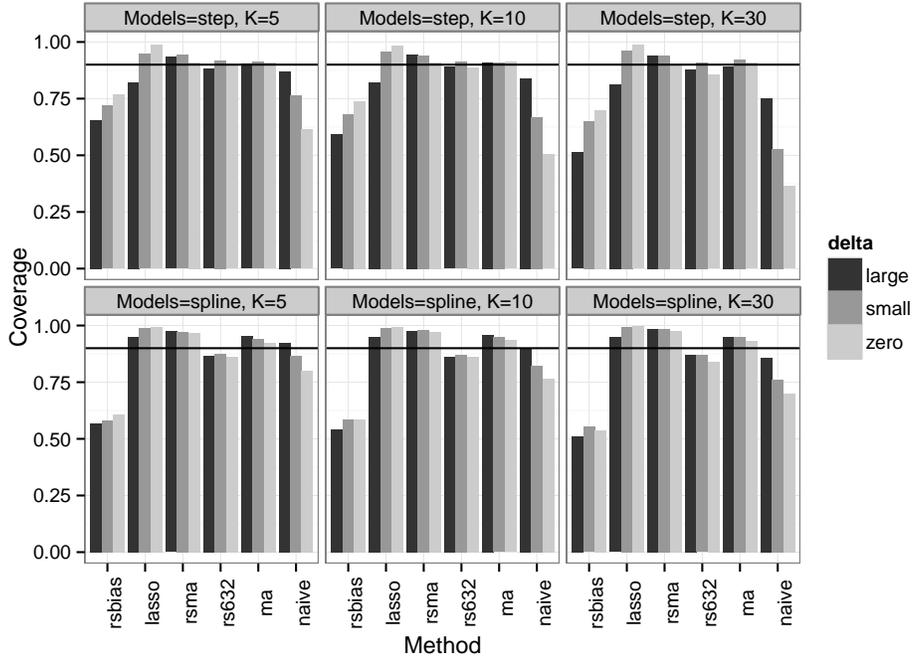

Figure 5: Average coverage of 90%-confidence intervals of estimators for the step-function (top row) and spline modeling approaches. Scenarios with differing number of covariates $K$ and effect sizes *delta* in the true subgroup are depicted. For all depicted scenarios $n = 50$ and $curve = step$.

$K$, $n$ and *delta* were varied.

These simulations aim to show the asymptotic behaviour of the estimates for the true subgroup as the evidence a subgroup effect gets stronger and stronger (or weaker). One specific question that these simulations tried to answer was, if the correct subgroup is known, how big does the evidence have to be, that the estimation method will give an estimate that is reasonably close to the naive (and in this case true) treatment effect. In short, when do the methods stop shrinking?

In addition we considered the situation that covariates are no longer generated independently, but are correlated with an AR(1)-correlation structure. Since correlation did not seem to influence the estimates, results for correlated covariates are not depicted here and can be found in the supplementary material.

For the simulations in this section the subgroup identification process was omitted and instead the subgroup $S = \{i \in \{1, ..., n\}|x_{i1} > median(x_{\cdot 1})\}$ was always chosen. The step function models were used for estimation and the true treatment effect curve also had the form of a step function. Unless one of the parameters was changed explicitly the sample size was $n = 500$ the overall average treatment effect was 0.29, which results in an overall power of 0.9, and the number of covariates was $K = 10$. Since a step function curve was used the true effect in the subgroup $S$ was 0.58 (see Table 3).

Results of these simulation runs are shown in Figure 6 for the three parameters $n$, $K$ and *delta*. Depicted are the averaged estimates over 5000 simulations. The sample size $n$ (Figure 6 (i)) seems to affect all estimates similarly, bringing them closer to the true effect as the sample size grows. In general all estimates apart from the Lasso seem to approach the true estimates as the sample size increases. The Lasso estimator seems to approach a value for the treatment effect, which lies a good amount below the true effect.

As it is visible from Figure 6 (ii) a higher number of covariates shrinks the estimates towards the overall effect. The ma-estimator and to a lesser amount the rsma- and rs632-estimators still remain close to the naive estimates even for a high number of covariates, as desired. The Lasso-estimator and especially the rsbias-estimator are more heavily affected by a high number of covariates. The rsbias-estimator even shrinks below the overall treatment effect.

In Figure 6 (iii) the effect size *delta* in the subgroup was modified. Similar to Figure 6 (i) all estimators except the Lasso approach the true effect as the effect size grows.

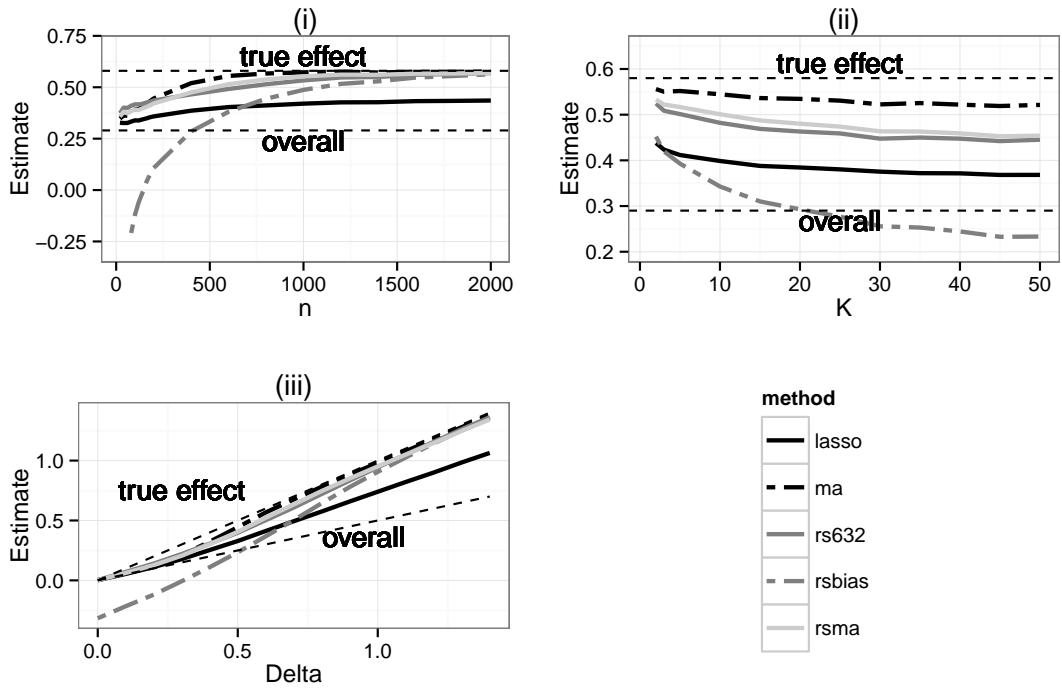

Figure 6: Influence of simulation parameters $n$ (i), $K$ (ii) and *delta* (iii) in the subgroup on different estimators for the subgroup $x_1 > median(x_1)$. The standard scenario is $n = 500$, $K = 10$, $theta = 0$ and a large effect size in the subgroup (0.58). The treatment effect curve has the form of a step function and the models using step functions are used.

# 6  Discussion

The purpose of this paper was to compare several methods for treatment effect estimation in subgroup analyses in a simulation study. We also considered the effect of categorizing covariates compared to modelling continuous covariates with splines.

One main conclusion to be drawn from the simulations in Section 4.2 is that naive estimation methods (ignoring model uncertainty/subgroup selection) will lead to inadequate inference: Treatment effects in selected subgroups will be over-estimated and confidence intervals will not have the nominal level. In addition we have shown that a few methods outperform the naive estimate in our scenarios: both in terms of bias and MSE of estimate but also in terms of confidence interval coverage, promising methods appear to be the model averaging (ma and rsma) approaches, the 0.632 bootstrap estimate (rs632) as well as the Lasso approach. The naive bias adjustment method (rsbias) based on directly estimating the bias caused by selection and subtracting it from the naive estimate, did not perform very well.

From the simulation results it seems that model averaging, resampling and Lasso are all viable candidates to adjust treatment effect estimates. In what follows we will compare these methods from a conceptual and computational perspective.

The adjustments induced by the model averaging approaches (ma, rsma) and the Lasso approach do not explicitly depend on the selection mechanism itself. These methods induce shrinkage in the treatment effect estimate from their underlying modelling assumptions. The resampling approaches (rsbias, rs632) depend on the specific selection mechanism, in the sense that different adjustments would be derived for different selection mechanisms. This also means that the ma, rsma and Lasso approach can be used to derived adjusted treatment effect estimates essentially for all subgroups $S$ one is interested in (this itself could also be utilized as a subgroup selection strategy). The resampling approaches can only make a statement about the actually selected subgroup. Finally the treatment effect estimates for a particular subgroup derived with the ma, rsma and Lasso approaches are only based on patients in that subgroup, while the adjustments in rsbias and rs632 are also based on patients not in the selected subgroup.

From a computational perspective the resampling methods are significantly more bur-densome, since subgroup identification has to be repeated in every bootstrap sample. For reference, estimation with resampling using 100 bootstrap samples took approximately 80 times longer than estimation with model averaging and approximately 20 times longer than estimation with lasso on our machine.

Even though the focus of our work was treatment effect estimation we also got some insight into the effect of categorizing covariates through our simulations. For determining the right covariate (defining the subgroup), it turned out that the step-function approach worked quite well compared to a spline-based approach (see Section 4.1). This is probably due to difficulty to estimate the splines model reliably with the relatively high variance (low signal to noise ratio) in the endpoint, even though a low dimensional basis was used (see Appendix A). Working with rank-transformed covariates improved the performance of the splines, in particular by reducing the variability at the end-points of the covariate space. P-splines (Eilers & Marx (1996)) or fractional polynomials (Royston & Sauerbrei (2004)), might further improve the situation, as the complexity of the model-fit is adaptively chosen by these methods (by using shrinkage or model selection). Of course, when it comes to estimation of the treatment effect curve $g(x)$ itself in an unweighted way the spline based approach might outperform the step-function approach, a metric we did not consider in this form in this paper. However, in our results naive estimation improved, when basing it upon splines.

For identification of subgroups we chose the group, which maximized standardized difference between treatment effect estimates (see Section 2.3). This implicitly penalizes towards groups of equal size. Alternatively one could also pick the subgroup to maximize the treatment effect difference (omitting the standardization by the standard error of the difference): The simulations from Sections 4.1 and 4.2 were also run with this identification rule. Results were however mostly unaffected by this change, noticeable was only a small improvement in correct identifications when using step functions.

# Acknowledgements

This work was supported by funding from the European Union's Horizon 2020 research and innovation programme under the Marie Sklodowska-Curie grant agreement No 633567 and by the Swiss State Secretariat for Education, Research and Innovation (SERI) under contract number 999754557 . The opinions expressed and arguments employed herein do not necessarily reflect the official views of the Swiss Government.

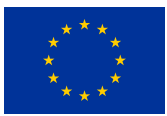
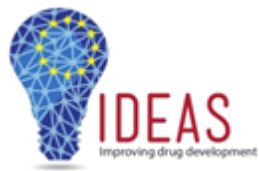

# Appendix A    Knot Selection Rules

We use quantiles to select the knots. We roughly follow the guidelines given by Harrell (2001). Compared to the original proposal less knots are selected to obtain less variable spline estimates. We base the number of knots on the number of patients in the treatment group and then use the same number of knots for the prognostic term $h$ and for the predictive term $g$. The knots used for different sample sizes in this simulation study are shown in Table A1.

Table A1: Number and placement of knots for natural cubic splines for different number of patients in the treatment arms, $x_{(i)}$ denotes the i-th ordered observation of $x$

| patients in trt | number of knots | placement of knots |
|:---:|:---:|:---:|
| $n \leq 10$ | 1 | median |
| $10 < n \leq 20$ | 2 | $x_{(5)}$, $x_{(n-4)}$ |
| $20 < n \leq 30$ | 3 | $x_{(5)}$, median, $x_{(n-4)}$ |
| $n \geq 30$ | 4 | $x_{(5)}$, 0.35-quantile, 0.65-quantile, $x_{(n-4)}$ |

# Appendix B  Obtaining Treatment Effect Curve Parameters

The overall treatment effect under model (14) is given by $\int_{-\infty}^{\infty} g(x) \cdot f_{N(0,1)}(x)dx$. Now we desire that a specific overall treatment effect holds for the population: $\int_{-\infty}^{\infty} g(x) \cdot f_{N(0,1)}(x)dx = \Delta(\Omega)$. An additional constraint is needed to obtain intercept and slope ($a$ and $b$) for all functions $g$. One possibility is to require that $\int_{0}^{\infty} g(x) \cdot f_{N(0,1)}(x)dx = 2\Delta(\Omega)$, i.e. the patients with positive covariate have an average effect, that is double the overall effect. For the step function $g_{step}$ this condition is fulfilled for $b_{step} = 2\Delta(\Omega)$. The other two functions can be written as $g(x) = a + b\tilde{g}(x)$, so that $a$ and $b$ can be moved out of the integral and $\int_{-\infty}^{\infty} \tilde{g}(x) \cdot f_{N(0,1)}(x)dx$ and $\int_{0}^{\infty} \tilde{g}(x) \cdot f_{N(0,1)}(x)dx$ can be precalculated so that $a$ and $b$ can be obtained by solving the corresponding linear system of two equations, as for both cases $\int_{-\infty}^{\infty} \tilde{g}(x) \cdot f_{N(0,1)}(x)dx = 0$, $a$ is equal to the overall effect, while one needs to solve for $b$. All solutions are given in Table 3.

## SUPPLEMENTARY MATERIAL

**Additional Simulation Results**  Simulation results for settings not depicted in the paper (pdf)

# Supplementary Material for: "Comparing Approaches to Treatment Effect Estimation for Subgroups in Clinical Trials"

# Contents

# S1 Additional Simulation Results

## S1.1 Subgroup Identification

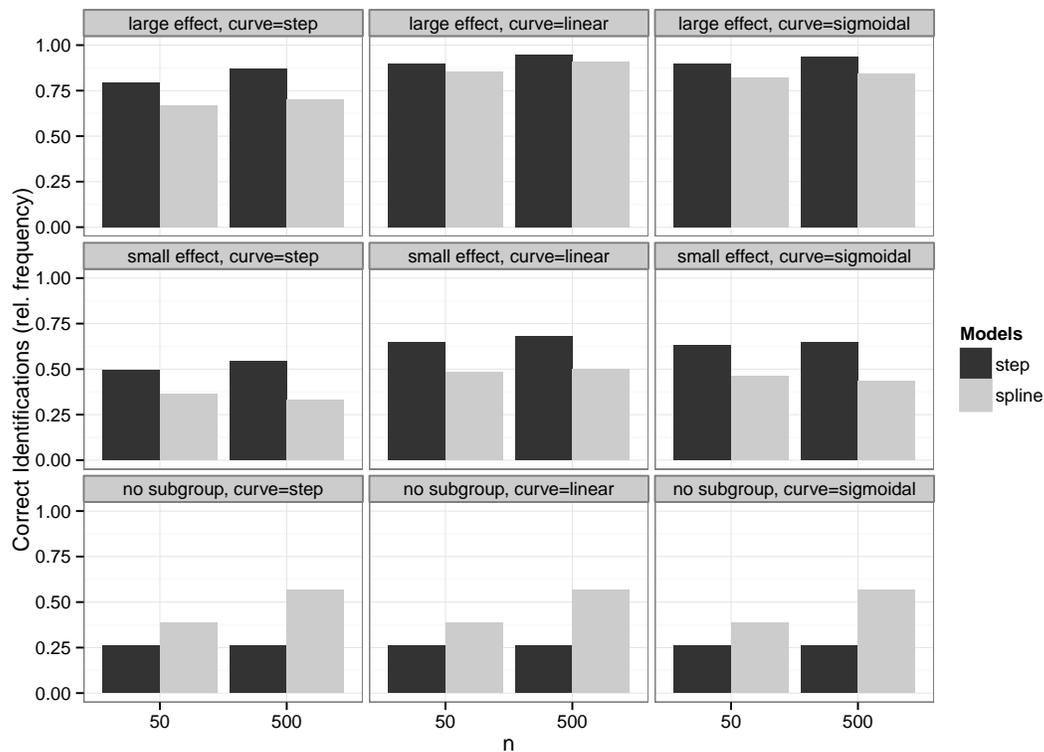

Figure S1: Relative frequency of correct identifications for the spline and step function modeling approaches. An identification is considered as correct, when there is a true interaction between the identified covariate and the treatment. In the case of no subgroup (last row) the correct decision is to not identify any subgroup. In all depicted scenarios the number of covariates was 5.

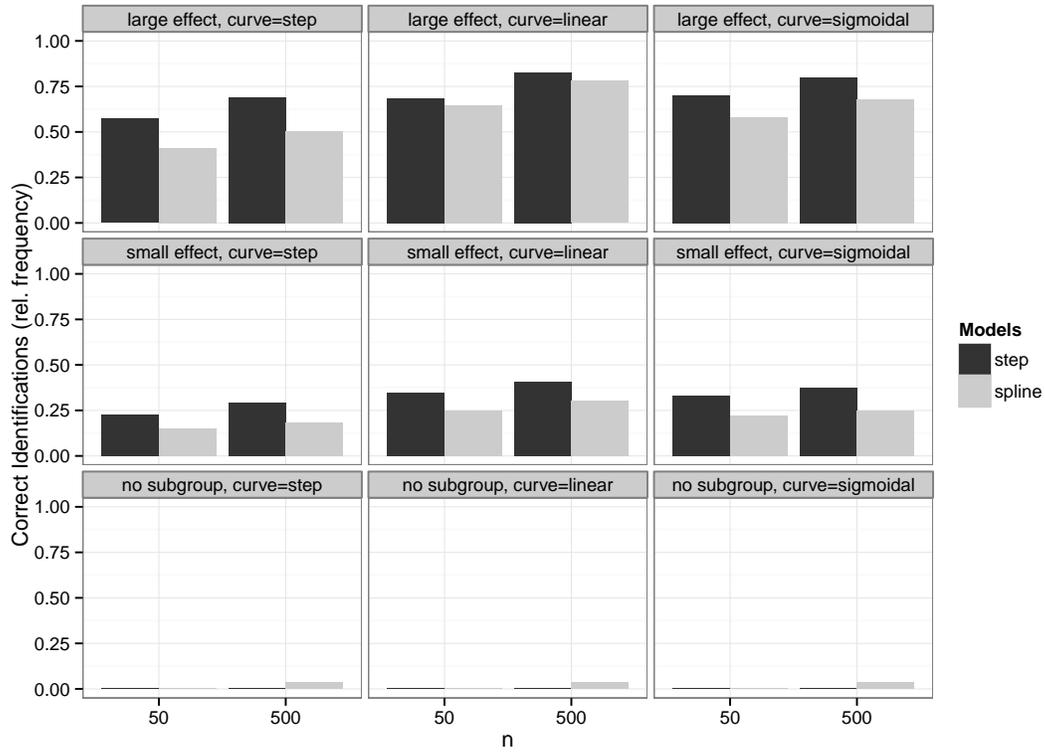

Figure S2: Relative frequency of correct identifications for the spline and step function modeling approaches. An identification is considered as correct, when there is a true interaction between the identified covariate and the treatment. In the case of no subgroup (last row) the correct decision is to not identify any subgroup. In all depicted scenarios the number of covariates was 30.

## S1.2 Effect Estimation in Subgroups

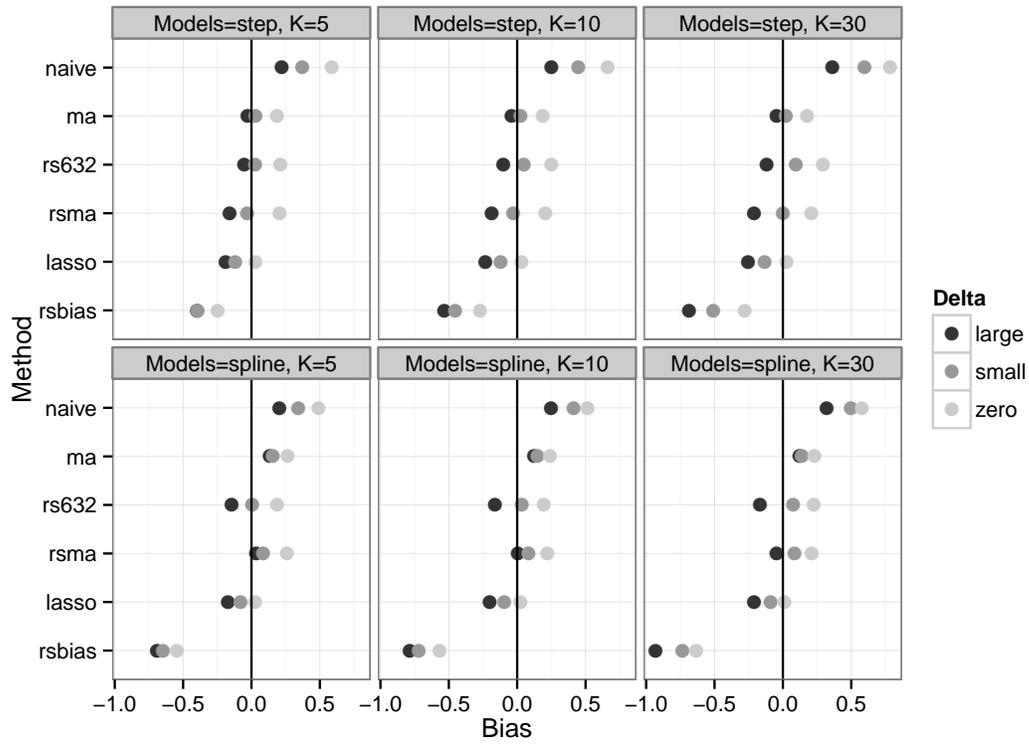

Figure S3: Median bias of estimators for the step-function (top row) and spline modeling approaches. Scenarios with differing number of covariates $K$ and effect sizes $Delta$ in the true subgroup are depicted. For all depicted scenarios $n = 50$ and $curve = linear$

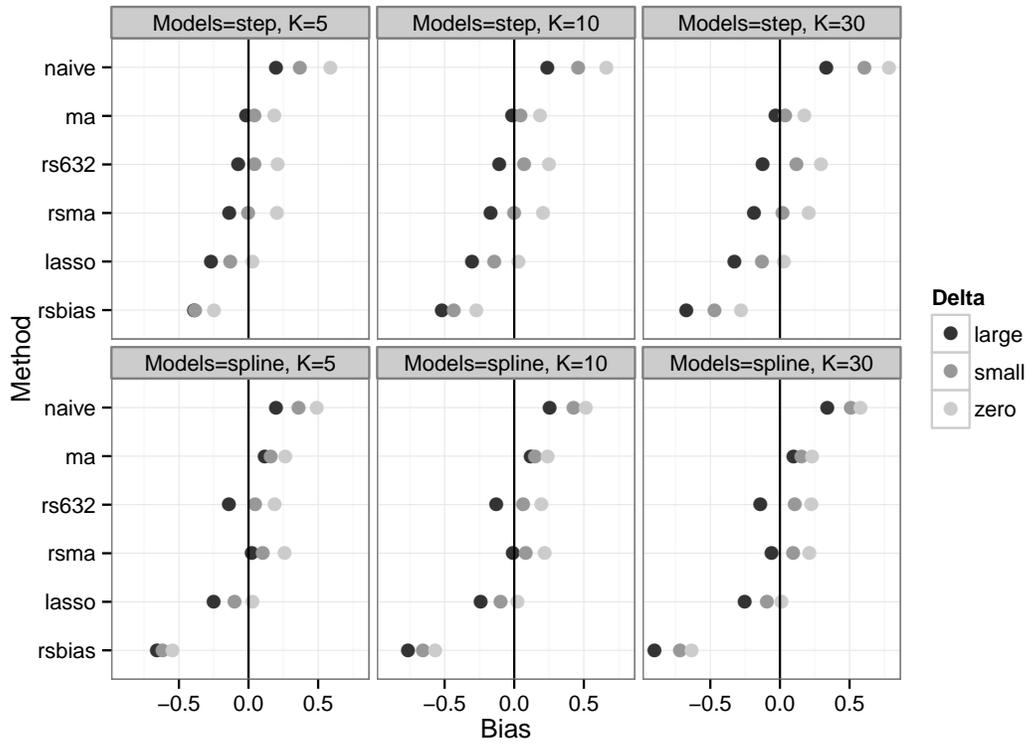

Figure S4: Median bias of estimators for the step-function (top row) and spline modeling approaches. Scenarios with differing number of covariates $K$ and effect sizes $Delta$ in the true subgroup are depicted. For all depicted scenarios $n = 50$ and $curve = sigmoidal$

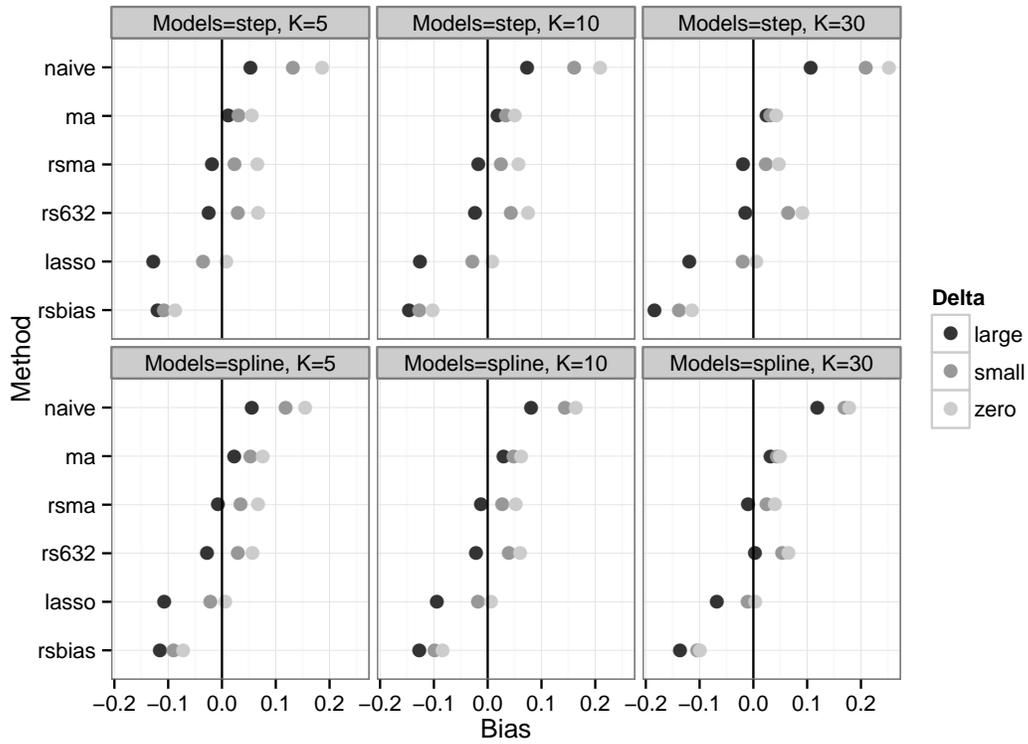

Figure S5: Median bias of estimators for the step-function (top row) and spline modeling approaches. Scenarios with differing number of covariates $K$ and effect sizes $Delta$ in the true subgroup are depicted. For all depicted scenarios $n = 500$ and $curve = step$

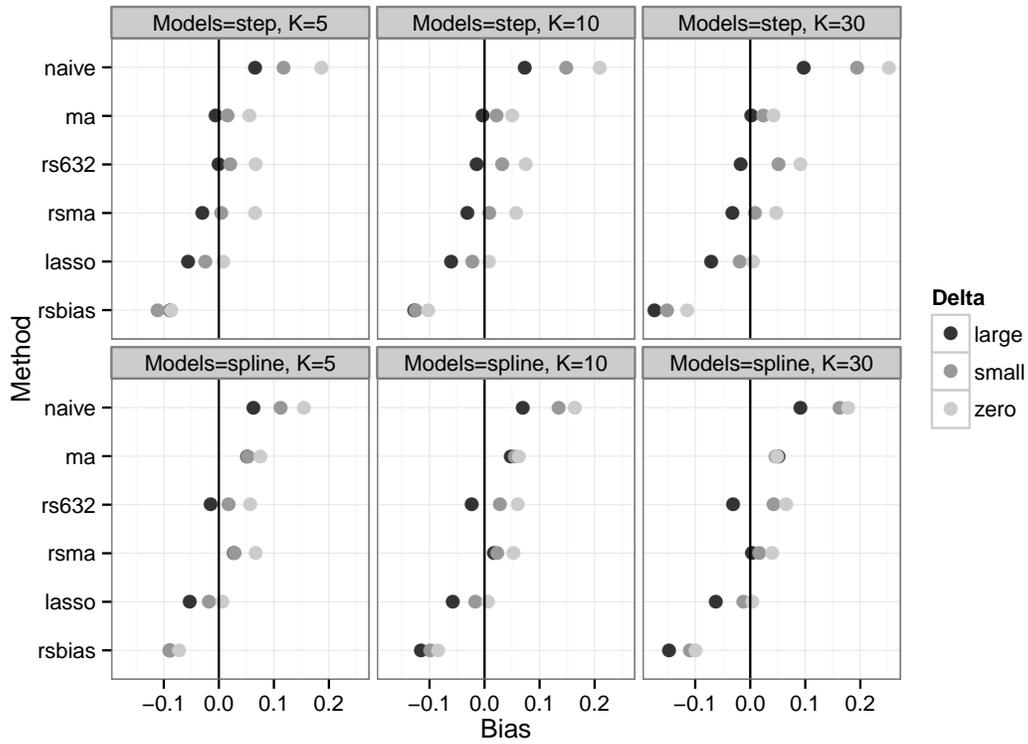

Figure S6: Median bias of estimators for the step-function (top row) and spline modeling approaches. Scenarios with differing number of covariates $K$ and effect sizes $Delta$ in the true subgroup are depicted. For all depicted scenarios $n = 500$ and $curve = linear$

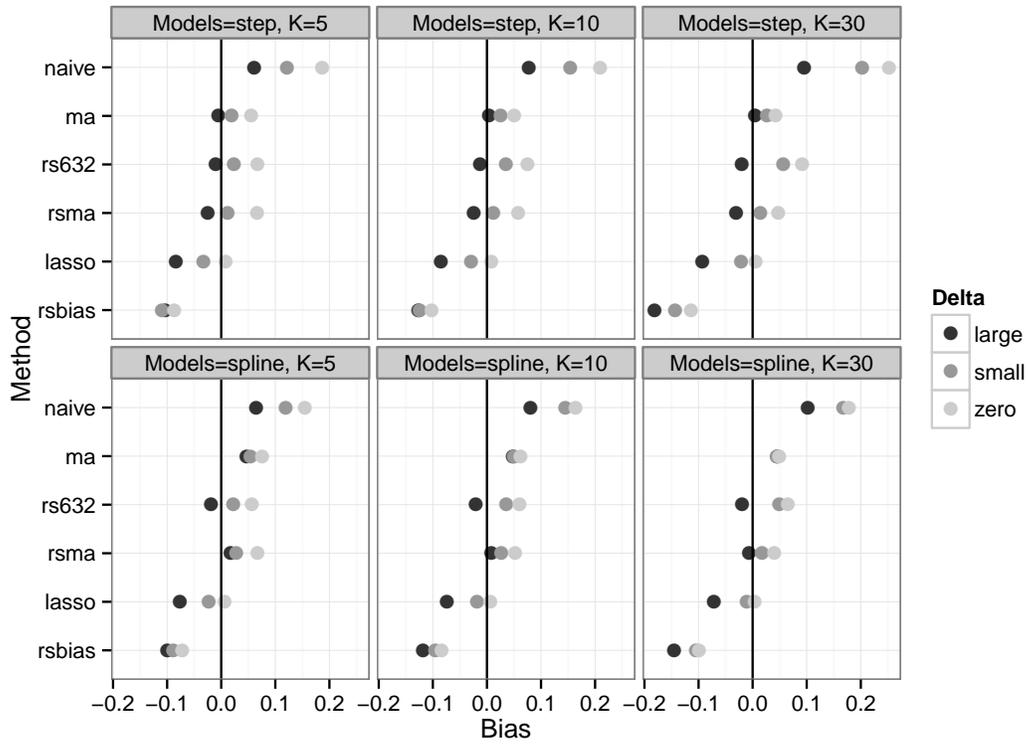

Figure S7: Median bias of estimators for the step-function (top row) and spline modeling approaches. Scenarios with differing number of covariates $K$ and effect sizes $Delta$ in the true subgroup are depicted. For all depicted scenarios $n = 500$ and $curve = sigmoidal$

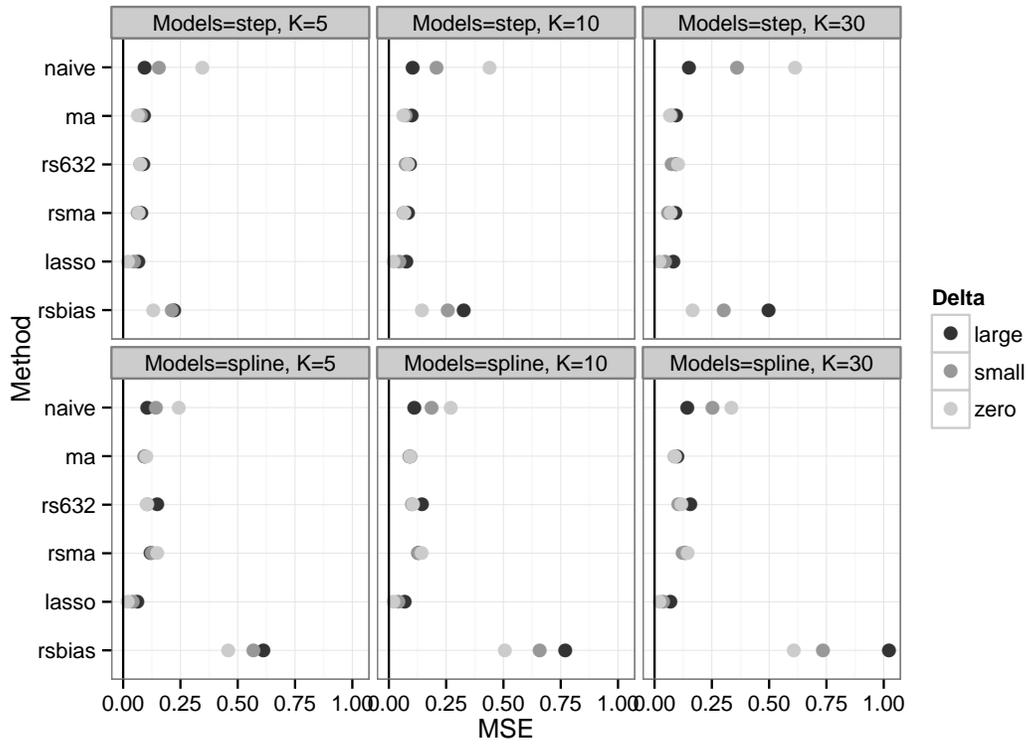

Figure S8: Median MSE of estimators for the step-function (top row) and spline modeling approaches. Scenarios with differing number of covariates $K$ and effect sizes $Delta$ in the true subgroup are depicted. For all depicted scenarios $n = 50$ and $curve = linear$

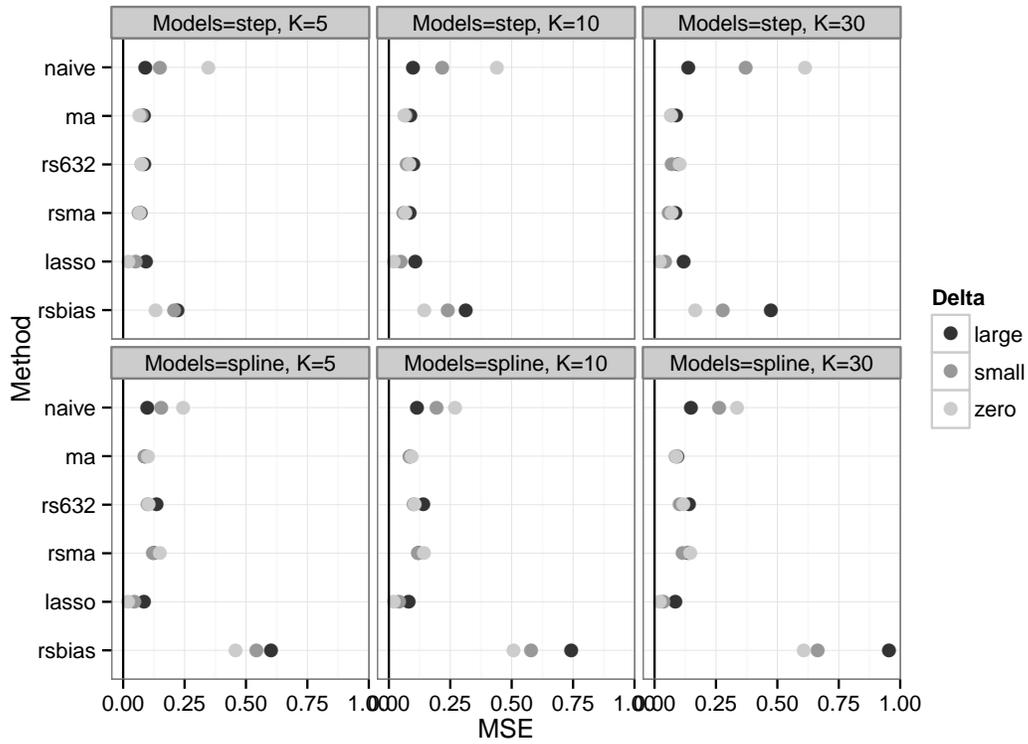

Figure S9: Median MSE of estimators for the step-function (top row) and spline modeling approaches. Scenarios with differing number of covariates $K$ and effect sizes $Delta$ in the true subgroup are depicted. For all depicted scenarios $n = 50$ and $curve = sigmoidal$

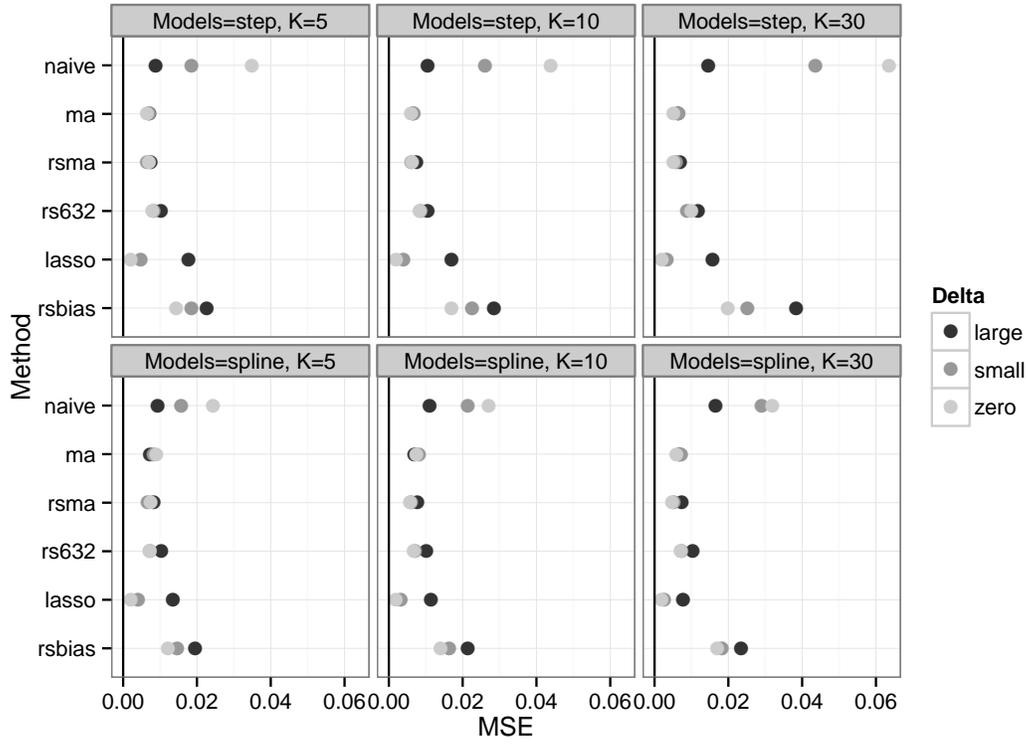

Figure S10: Median MSE of estimators for the step-function (top row) and spline modeling approaches. Scenarios with differing number of covariates $K$ and effect sizes $Delta$ in the true subgroup are depicted. For all depicted scenarios $n = 500$ and $curve = step$

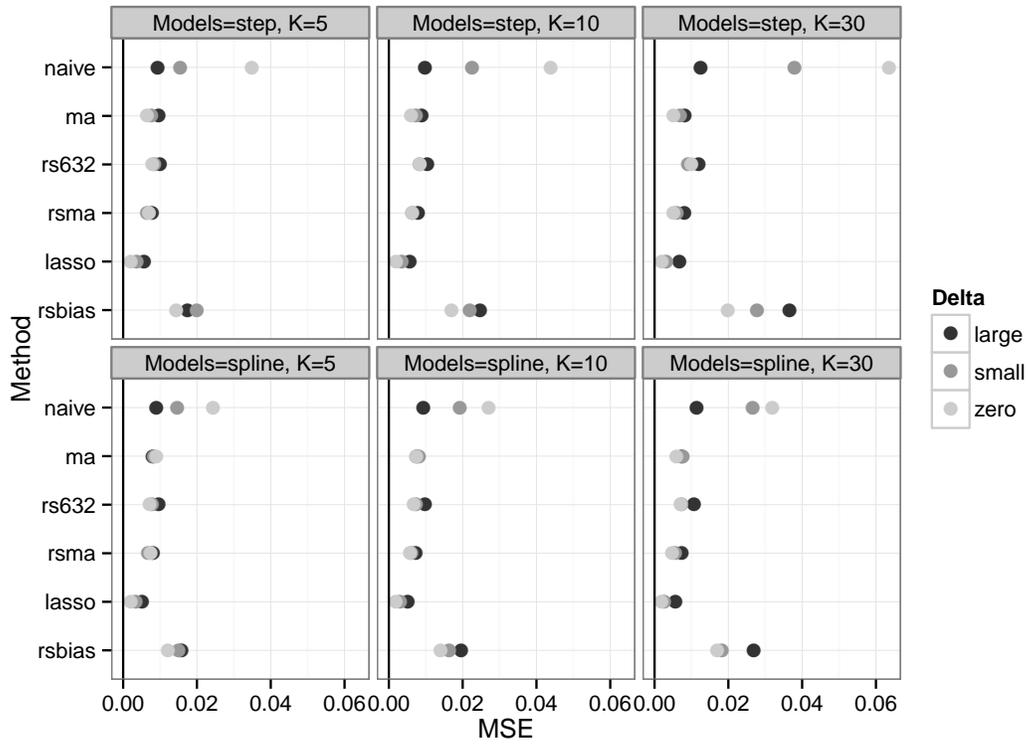

Figure S11: Median MSE of estimators for the step-function (top row) and spline modeling approaches. Scenarios with differing number of covariates $K$ and effect sizes $Delta$ in the true subgroup are depicted. For all depicted scenarios $n = 500$ and $curve = linear$

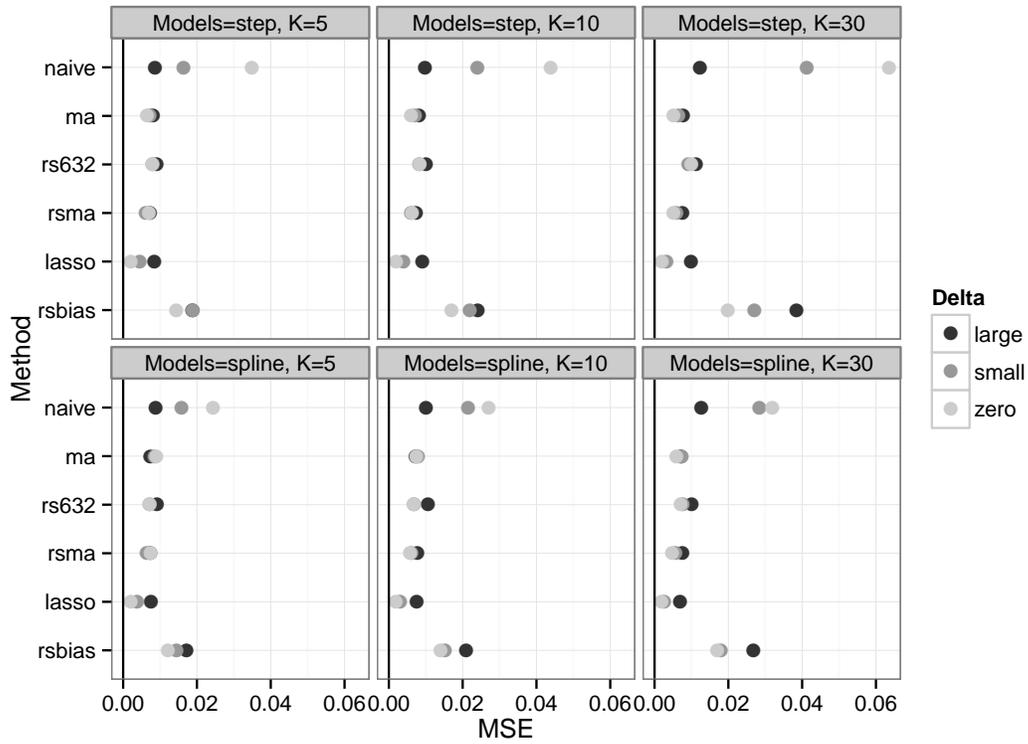

Figure S12: Median MSE of estimators for the step-function (top row) and spline modeling approaches. Scenarios with differing number of covariates $K$ and effect sizes $Delta$ in the true subgroup are depicted. For all depicted scenarios $n = 500$ and $curve = sigmoidal$

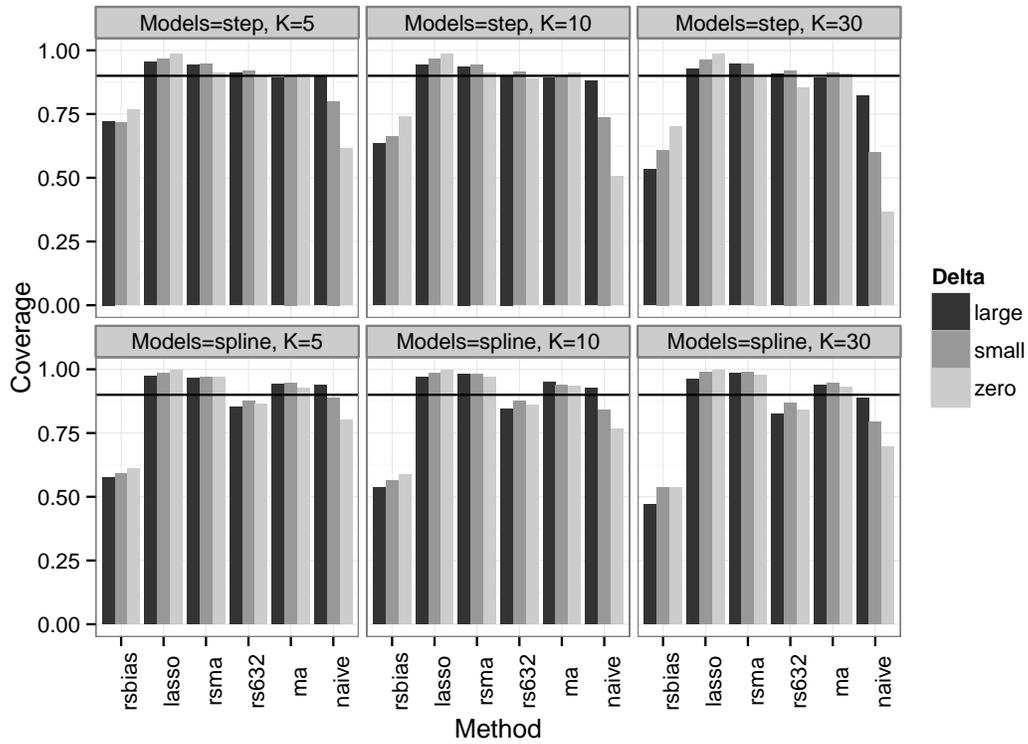

Figure S13: Average coverage of 90%-confidence intervals of estimators for the step-function (top row) and spline modeling approaches. Scenarios with differing number of covariates $K$ and effect sizes $Delta$ in the true subgroup are depicted. For all depicted scenarios $n = 50$ and $curve = linear$

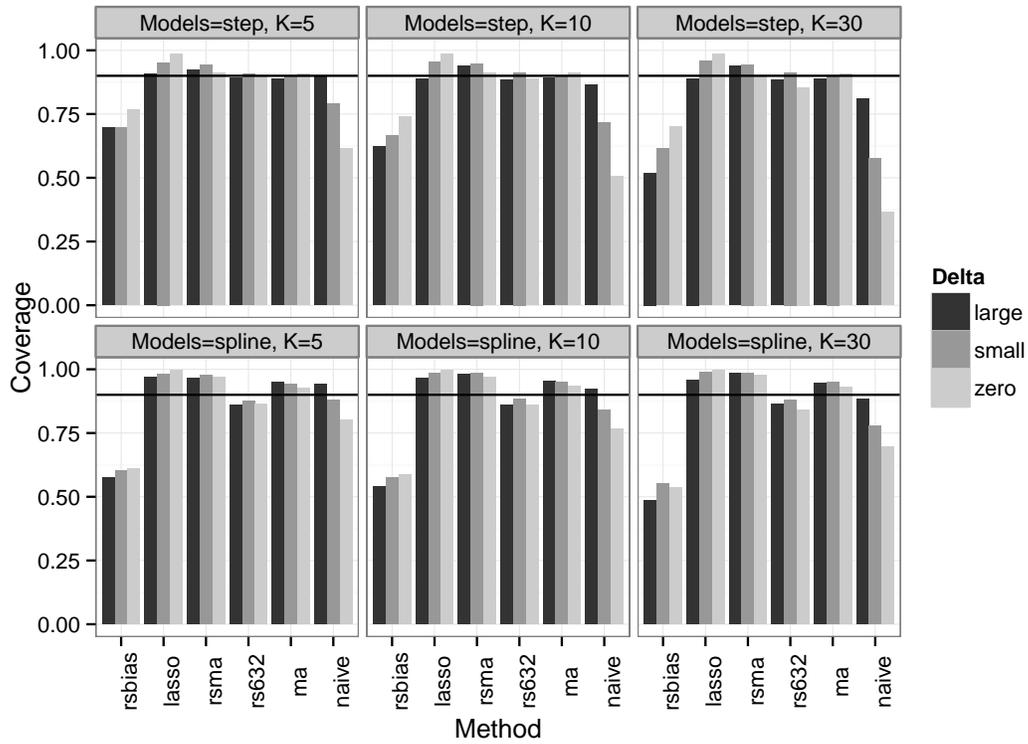

Figure S14: Average coverage of 90%-confidence intervals of estimators for the step-function (top row) and spline modeling approaches. Scenarios with differing number of covariates $K$ and effect sizes $Delta$ in the true subgroup are depicted. For all depicted scenarios $n = 50$ and $curve = sigmoidal$

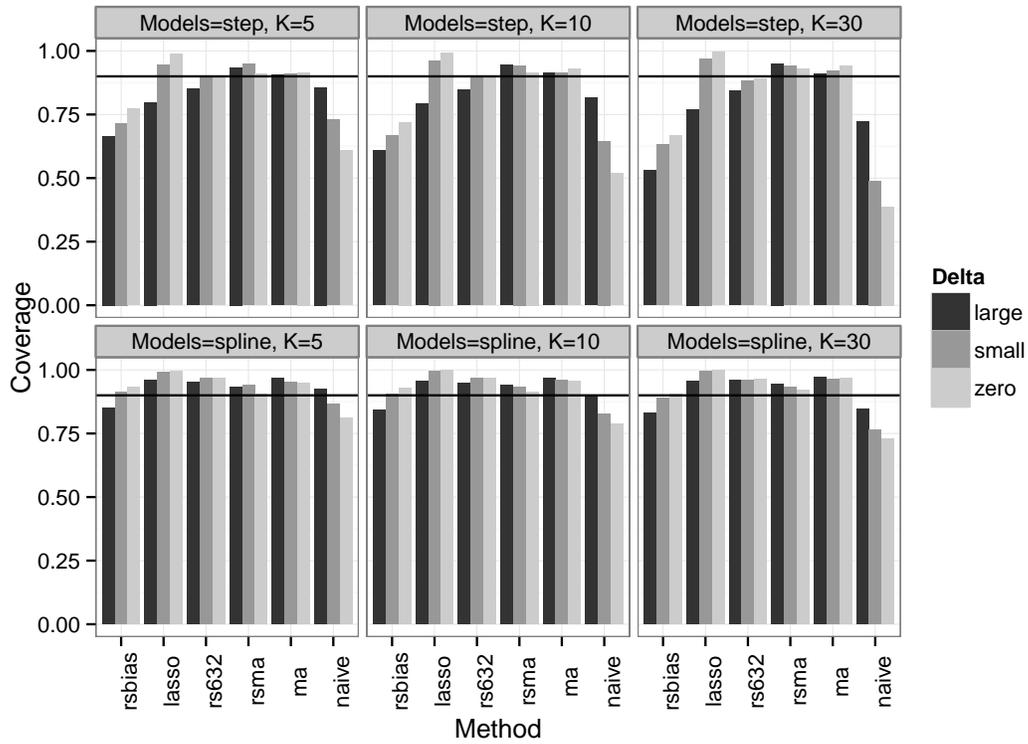

Figure S15: Average coverage of 90%-confidence intervals of estimators for the step-function (top row) and spline modeling approaches. Scenarios with differing number of covariates $K$ and effect sizes $Delta$ in the true subgroup are depicted. For all depicted scenarios $n = 500$ and $curve = step$

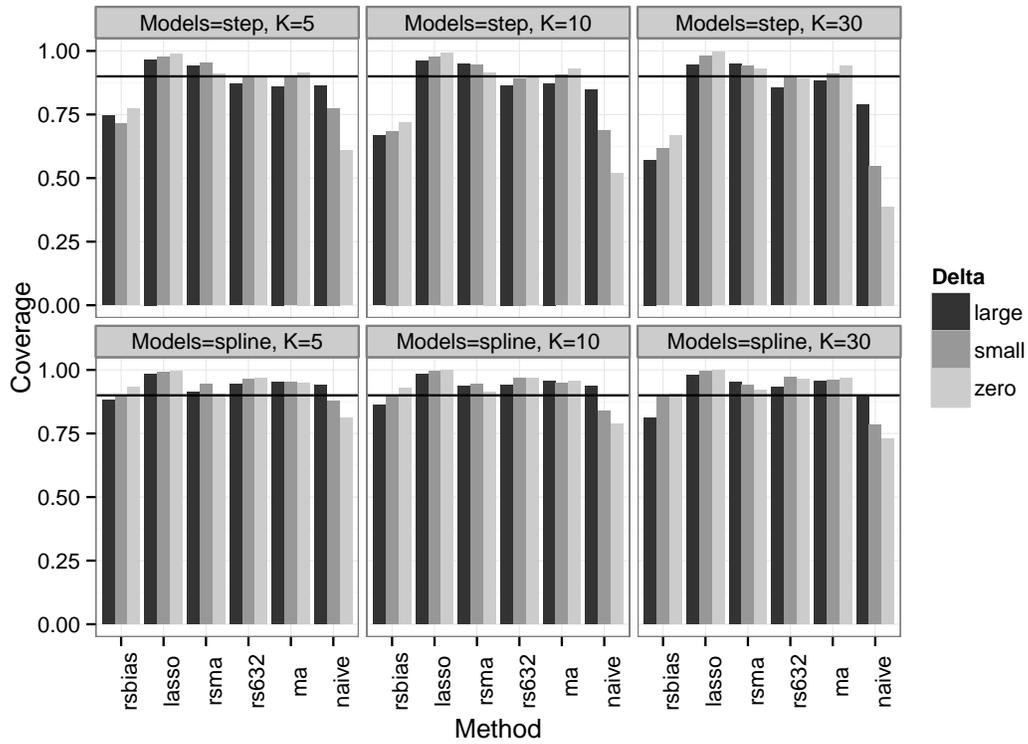

Figure S16: Average coverage of 90%-confidence intervals of estimators for the step-function (top row) and spline modeling approaches. Scenarios with differing number of covariates $K$ and effect sizes $Delta$ in the true subgroup are depicted. For all depicted scenarios $n = 500$ and $curve = linear$

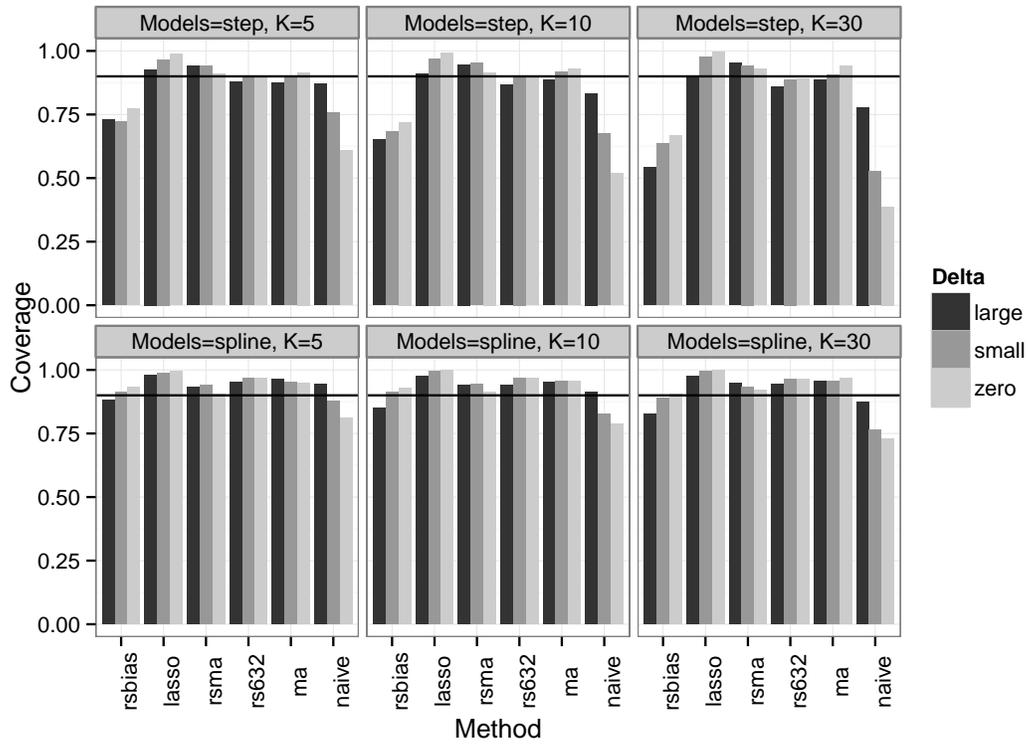

Figure S17: Average coverage of 90%-confidence intervals of estimators for the step-function (top row) and spline modeling approaches. Scenarios with differing number of covariates $K$ and effect sizes $Delta$ in the true subgroup are depicted. For all depicted scenarios $n = 500$ and $curve = sigmoidal$

## S1.3 Influence of Simulation Parameter: Correlation

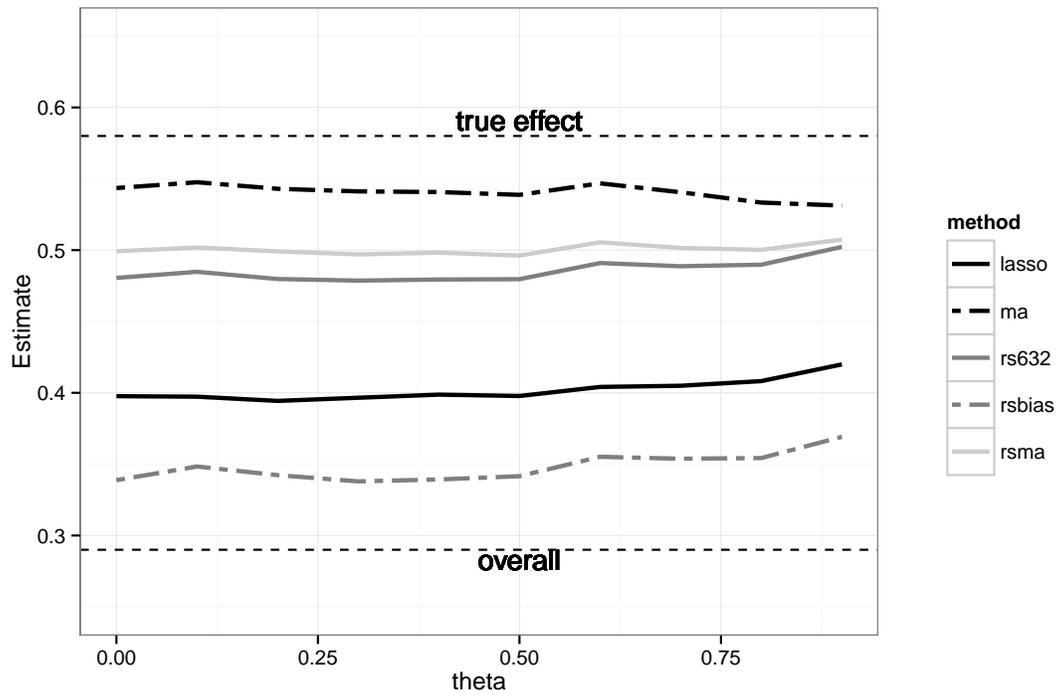

Figure S18: Influence of correlation between covariates $theta$ on different estimators for the subgroup $x_1 > median(x_1)$. The standard scenario is $n = 500$, $K = 10$, $theta = 0$ and a large effect size in the subgroup (0.58). The treatment effect curve has the form of a step-function and the models with step-functions are used.